\documentclass[twoside,twocolumn]{IEEEtran}

\usepackage{graphicx}
\usepackage{dcolumn}
\usepackage{array}
\usepackage{amsmath}
\usepackage{fancyheadings}
\usepackage{subfigure}
\usepackage{algorithm}
\usepackage{algorithmic}
\usepackage{verbatim}
\usepackage{tabularx}

\newcolumntype{d}[1]{D{.}{.}{#1}}
\newcolumntype{i}{D{.}{}{0}}

\begin{document}
\pagestyle{empty}
\date{}
\title{Analytical Framework for Optimizing Weighted Average Download Time in Peer-to-Peer Networks}
\author{Bike Xie, Mihaela van der Schaar and Richard D. Wesel\\
Department of Electrical Engineering, University of California, Los
Angeles, CA 90095-1594\\Email:
xbk@ee.ucla.edu, mihaela@ee.ucla.edu, wesel@ee.ucla.edu
\thanks{}}

\maketitle \thispagestyle{empty}

\begin{abstract}
This paper proposes an analytical framework for peer-to-peer (P2P)
networks and introduces schemes for building P2P networks to
approach the minimum weighted average download time (WADT). In the
considered P2P framework, the server, which has the information of
all the download bandwidths and upload bandwidths of the peers,
minimizes the weighted average download time by determining the
optimal transmission rate from the server to the peers and from the
peers to the other peers. This paper first defines the static P2P
network, the hierarchical P2P network and the strictly hierarchical
P2P network. Any static P2P network can be decomposed into an
equivalent network of sub-peers that is strictly hierarchical. This
paper shows that convex optimization can minimize the WADT for P2P
networks by equivalently minimizing the WADT for strictly
hierarchical networks of sub-peers. This paper then gives an upper
bound for minimizing WADT by constructing a hierarchical P2P
network, and lower bound by weakening the constraints of the convex
problem. Both the upper bound and the lower bound are very tight.
This paper also provides several suboptimal solutions for minimizing
the WADT for strictly hierarchical networks, in which peer selection
algorithms and chunk selection algorithm can be locally designed.
\end{abstract}


\section{Introduction}

P2P applications (e.g, \cite{Napster}, \cite{Gnutella},
\cite{KaZaA}) are increasingly popular and represent a large
majority of the traffic currently transmitted over the Internet. A
unique feature of P2P networks is their flexible and distributed
nature, where each peer can act as both server and client
\cite{AND04}. Hence, P2P networks provide a cost-effective and
easily deployable framework for disseminating large files without
relying on a centralized infrastructure \cite{Liu07}. These features
of P2P networks have made them popular for a variety of broadcasting
and file-distribution applications \cite{Liu07} \cite{Zhang05}
\cite{Pai05} \cite{Li04} \cite{Xiang04}. Specifically, chunk-based
and data-driven P2P broadcasting systems such as CoolStreaming
\cite{Zhang05}, Overcast \cite{Overcast} and Chainsaw \cite{Pai05}
have been developed, which adopt pull-based techniques
\cite{Zhang05}, \cite{Pai05}. In these P2P systems, the peers
possess several chunks and these chunks are shared by peers that are
interested in the same content. An important problem in such P2P
systems is how to transmit the chunks to the various peers and
create reliable and efficient connections between peers. For this,
various approaches have been proposed including tree-based and
data-driven approaches (e.g. \cite{Li04} \cite{Chu04} \cite{DES01}
\cite{Jiang03} \cite{Cui04} \cite{PAD03} \cite{PAD02}).

Besides these practical approaches, some research has begun to
analyze P2P networks from a theoretic perspective to quantify the
achievable performance. The performance, scalability and robustness
of P2P networks using network coding are studied in \cite{Chou04}
\cite{ACE05}. In these investigations, each peer in a P2P network
randomly chooses several peers including the server as its parents,
and also transmits to its children a random linear combination of
all packets the peer has received. Network coding, working as a
perfect chunk selection algorithm, guarantees every packet
transmitted in a P2P network has new information for its receiver,
which makes elegant theoretical analysis possible. Other research
studies the steady-state behavior of P2P networks with homogenous
peers using fluid models \cite{Qiu04} \cite{Ge03} \cite{CLE04}. Most
papers providing theoretical analysis for P2P networks assume
dynamic systems with homogenous peers.

This paper establishes an analytical framework for optimizing
weighted average download time (WADT) for P2P networks with
heterogeneous peers, i.e., peers with different download bandwidths
and upload bandwidths. This paper focuses on static P2P networks
with a single server and a fixed number of peers. In other words, no
peer leaves or joins the P2P network. In the scheme of building the
P2P network, the server collects all the download bandwidths and
upload bandwidths of the peers, and minimizes the WADT by
determining the optimal transmission rate from the server or any
peer to any other peer. A dynamic P2P system can also be modeled as
a sequence of static P2P systems. Therefore, this study of static
P2P networks is the first step to a complete analytical framework.
We leave the study of dynamic P2P networks with heterogeneous peers
for future work.

A static P2P network is a directed graph which has one root, the
server, and at least one directed path from the root to each of the
other nodes, which are peers. In a P2P network, peers are placed
into levels according to the topological distances between these
peers and the server. A peer is in level $K$ if the length of the
longest directed acyclic path from the server to the peer is $K$. A
hierarchical P2P network is a P2P network in which each peer can
only download from peers in the lower levels and upload to peers in
the higher levels. Peers in the same level cannot download or upload
to each other. A strictly hierarchical P2P network is a P2P network
in which each peer in level $K$ can only download from peers in
level $K-1$ and upload to peers in level $K+1$.

This paper shows that any static P2P network can be decomposed into
an equivalent network of sub-peers that is strictly hierarchical.
Therefore, convex optimization can minimize the WADT for P2P
networks by equivalently minimizing the WADT for strictly
hierarchical networks of sub-peers. This paper then gives an
achievable upper bound for minimizing WADT by constructing a
hierarchical P2P network, and lower bound by weakening the
constraints of the convex problem. Both the upper bound and the
lower bound are very tight.

The strictly hierarchical P2P network is practical for protocol
design because peer selection algorithms and chunk selection
algorithms can be locally designed level by level instead of
globally designed. Minimizing the WADT for strictly hierarchical
networks is a 0-1 convex optimization problem. However, if we have
assigned all peers each to a level, then the global bandwidth
allocation problem decomposes into local bandwidth allocation
problems at each level, which have water-filling solutions. Several
suboptimal peer assignment algorithms are provided and simulated.
Some of these suboptimal but practical schemes can be used for
content distribution systems, e.g. Overcast \cite{Overcast}.

This paper is organized as follows. In Section \ref{sec:setup},
definitions and notation for P2P networks are introduced. Section
\ref{sec:taxonomy} provides and discusses a taxonomy of overlay
networks. In Section \ref{sec:WADT}, the problem of minimizing the
weighted average download time is formulated and solved. Section
\ref{sec:results} presents the simulation results. Section
\ref{sec:conclusions} presents the conclusions.

\section{Setup and Problem Definition}
\label{sec:setup}

Consider a scenario where millions of peers would like to download
content from a server in the Internet. The server has sufficient
bandwidth to serve tens or hundreds of peers, but not millions. In
the absence of IP multicast, one solution is to form the server and
the peers into a P2P overlay network and distribute the content
using application layer multicast \cite{Chou04} \cite{Chu00}. In
this scenario, the content in the server is partitioned into chunks.
Peers not only download chunks from the server and other peers but
also upload to some other peers that are interested in the content.

This paper focuses on content distribution applications (e.g,
BitTorrent, Overcast \cite{Overcast}) in which peers are only
interested in content at full fidelity, even if it means that the
content does not become available to all peers at the same time. The
key issue for these P2P applications is to minimize download times
for peers. Since it usually takes several hours or days for a peer
to download content in full fidelity, our work is less concerned
with interactive response times and transmission delays in buffers
and in the network.

This paper studies a scheme to minimize the weighted average
download time for a static P2P network. In this scheme, the server
first collects the information of peers' weights, download
bandwidths, and upload bandwidths and then performs a centralized
optimization to find the best static P2P network, i.e., the optimal
transmission rates of the transmission flows from the server or any
peer to any other peer to minimize the weighted average download
time. The server passes the optimal solution to the peers, and the
peers build the connections according to the optimal solution. The
rest of the paper will focus on the centralized optimization
algorithm for determining the optimal static P2P network.

In a static P2P network, the server with bandwidth $S$ has a file,
whose size is 1 unit for simplicity. There are $N$ peers who want to
share the file in the network. Each peer has download bandwidth
$d_i$ and upload bandwidth $u_i$, for $i=1,2,\cdots,N$. These
download bandwidths and upload bandwidths are usually determined at
the application layer instead of the physical layer because an
Internet user can have several downloading tasks and these tasks
share the physical download and upload bandwidth of the user.

It is reasonable to assume that $d_i \geq u_i$ for each $1\leq i
\leq N$. For the case of $d_i < u_i$ for peer $i$, we just use the
part of the upload bandwidth which is the same as the download
bandwidth and leave the rest of the upload bandwidth.

\begin{figure}
  \centering
  \includegraphics[width=0.3\textwidth]{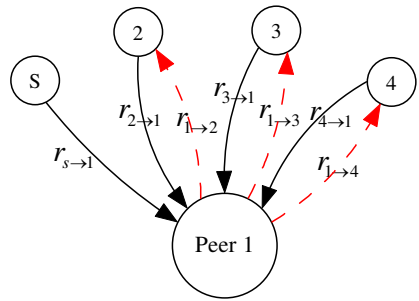}
  \caption{The peer model}\label{fig:peermodel}
\end{figure}

Denote the transmission rate from the server to peer $j$ as
$r_{s\rightarrow j}$ and the transmission rate from peer $i$ to peer
$j$ as $r_{i \rightarrow j}$. The total download rate of peer $j$,
denoted as $r_{j}$, is the summation of $r_{s\rightarrow j}$ and
$r_{i \rightarrow j}$ for all $i \neq j$. Since the total download
rate is constrained by the download bandwidth, we have $r_j =
r_{s\rightarrow j} + \sum_{i \neq j} r_{i \rightarrow j} \leq d_j$
for all $j=1,\cdots,N$. Since the total upload rate is constrained
by the upload bandwidth, we also have $\sum_{i \neq j} r_{j
\rightarrow i} \leq u_j$ for all $j=1,\cdots,N$. One example of the
peer model is shown in Fig.~\ref{fig:peermodel}. The download
bandwidth and upload bandwidth of Peer 1 are $d_1$ and $u_1$
respectively. Thus, the total download rate $r_{s\rightarrow 1} +
\sum_{i=2}^{4} r_{i \rightarrow 1}$ is less than or equal to $d_1$.
The total upload rate $\sum_{i=2}^{4} r_{1 \rightarrow i}$ is less
than or equal to $u_1$.

Since our work is less concerned with interactive response times and
transmission delays in buffers and in the network, the download time
for each peer is dominated by $1/r_i$ and the weighted average
download time is $\sum_{i=1}^N w_i /r_i$ where $w_i \geq 0$ is the
weight of peer $i$.

In a P2P network, if peer $i$ forwards content to peer $j$, then
peer $i$ is a parent of peer $j$ and peer $j$ is a child of peer
$i$. Each node can have several parents and several children. A
primary goal of this paper is to minimize the WADT computed as
$\sum_{i=1}^{N} w_i/r_i$. In the WADT computation $r_i$ could refer
to the actual download rate $r_i^{(a)}$ or the budgeted download
rate $r_i^{(b)}$. In general, $r_i^{(a)} \leq r_i^{(b)}$ because the
parents of a peer might not always have new content for sharing. The
network coding strategy can be used as a perfect chunk selection
algorithm to guarantee that a parent always has new content for its
children \cite{Chou04}. This paper uses the network coding strategy
as a chunk selection algorithm in P2P networks and so $r_i^{(a)} =
r_i^{(b)}$. Therefore, in the rest of the paper, we use $r_i$ for
both budgeted download rate and actual download rate.

\section{Taxonomy of Overlay Networks}
\label{sec:taxonomy}

This section studies graph structures of P2P overlay networks.

\newtheorem{P2P}{Definition}
\begin{P2P}\label{definition:P2P}
\textbf{static P2P network}: A static P2P network is a directed
graph which has one root and at least one directed path from the
root to each of the other nodes.
\end{P2P}

The root node is the server which has the content to share. All the
other nodes are the peers that are interested in the content. In a
P2P network, peers are placed into levels according to the
topological distances between these peers and the server.

\newtheorem{level}[P2P]{Definition}
\begin{level}\label{definition:level}
\textbf{level of peer}: A peer/node(use peer or node?) is in level
$K$ if the length of the longest directed acyclic path from the
server to the peer is $K$. The server is defined in level $0$.
\end{level}

\newtheorem{levelgraph}[P2P]{Definition}
\begin{levelgraph}\label{definition:levelgraph}
\textbf{level of P2P network}: A P2P network has level $K$ is the
maximum level of peers is $K$.
\end{levelgraph}

\subsection{The hierarchical P2P network}
\label{sec:HP2P}

\newtheorem{HP2P}[P2P]{Definition}
\begin{HP2P}\label{definition:HP2P}
\textbf{hierarchical P2P network}: A hierarchical P2P network is a
P2P network in which each peer can only download from peers in the
lower levels and upload to peers in the higher levels.
\end{HP2P}

\begin{figure}
  \centering
  \includegraphics[width=0.35\textwidth]{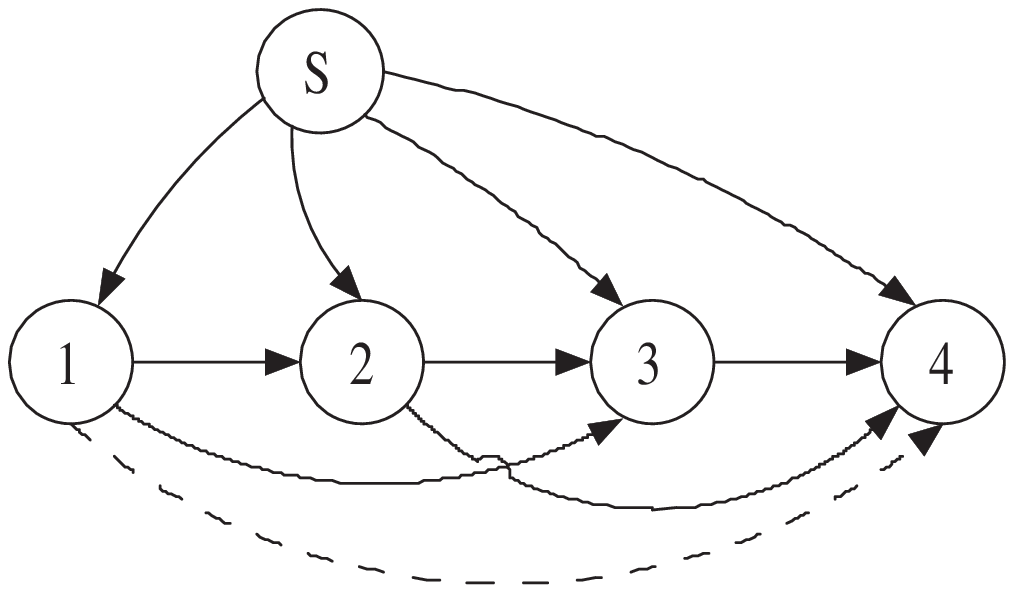}
  \caption{A hierarchical P2P network with 4 peers}\label{fig:HP2P}
\end{figure}

Fig.~\ref{fig:HP2P} shows a hierarchical P2P network with 4 peers,
in which peer $i$ is in level $i$. Note that in this example, each
peer downloads from all peers in the lower levels and uploads to all
peers in the higher levels. However, by definition, peers are not
required to download from all peers in the lower levels or upload
to/from? all peers in the higher levels. For example, the network
shown in Fig.~\ref{fig:HP2P} will still be a hierarchical P2P
network if the edge $1 \rightarrow 4$ vanishes. However, peers are
not allowed to download from any peer in a higher level or upload
to/from? any peer in a lower level. Also peers in the same level
cannot download or upload to each other.

\newtheorem{HP2PnoCycle}{Lemma}
\begin{HP2PnoCycle}\label{theorem:HP2PnoCycle}
A hierarchical P2P network contains no directed cycle.
\end{HP2PnoCycle}

\emph{Proof}:(by contraposition) Suppose there is a directed cycle
$A_1 \rightarrow A_2 \rightarrow \cdots \rightarrow A_n \rightarrow
A_1$, then some $A_i \rightarrow A_{i+1}$ or $A_n \rightarrow A_1$
violates the requirement that peers in a hierarchical P2P network
cannot upload to a peer in a lower level. Therefore, the network
containing the directed cycle cannot be a hierarchical P2P network.
Q.E.D.

\newtheorem{noCycleHP2P}[HP2PnoCycle]{Lemma}
\begin{noCycleHP2P}\label{theorem:noCycleHP2P}
A directed acyclic P2P network is a hierarchical P2P network.
\end{noCycleHP2P}

\emph{Proof}:(by contraposition) Suppose a P2P network is not
hierarchical, i.e., there exits a node $A$ in level $m$ and a node
$B$ in level $n$ such that $m \geq n$ and $A \rightarrow B$. Let $S
\rightarrow A_1 \rightarrow \cdots \rightarrow A_{m-1} \rightarrow
A$ be the longest directed acyclic path from $S$ to $A$ and $S
\rightarrow B_1 \rightarrow \cdots \rightarrow B_{n-1} \rightarrow
B$ be the longest directed acyclic path from $S$ to $B$. Since $S
\rightarrow A_1 \rightarrow \cdots \rightarrow A_{m-1} \rightarrow A
\rightarrow B$ is a directed path from $S$ to $B$ with length $m+1 >
n$, this path must contain a directed cycle, and hence, the P2P
network is not a directed acyclic graph. Therefore, a directed
acyclic P2P network is a hierarchical P2P network. Q.E.D.

\newtheorem{HP2P=NoCycle}{Theorem}
\begin{HP2P=NoCycle}\label{theorem:HP2P=NoCycle}
The set of all hierarchical P2P networks is the set of all directed
acyclic P2P networks.
\end{HP2P=NoCycle}

It is a direct consequence of Lemma \ref{theorem:HP2PnoCycle} and
Lemma \ref{theorem:noCycleHP2P}.

\subsection{The strictly hierarchical P2P network}
\label{sec:SHP2P}

\newtheorem{SHP2P}[P2P]{Definiation}
\begin{SHP2P}\label{definition:SHP2P}
\textbf{strictly hierarchical P2P network}: A strictly hierarchical
P2P network is a P2P network in which each peer in level $K$ can
only download from peers in level $K-1$ and upload to peers in level
$K+1$.
\end{SHP2P}

\begin{figure}
  \centering
  \includegraphics[width=0.25\textwidth]{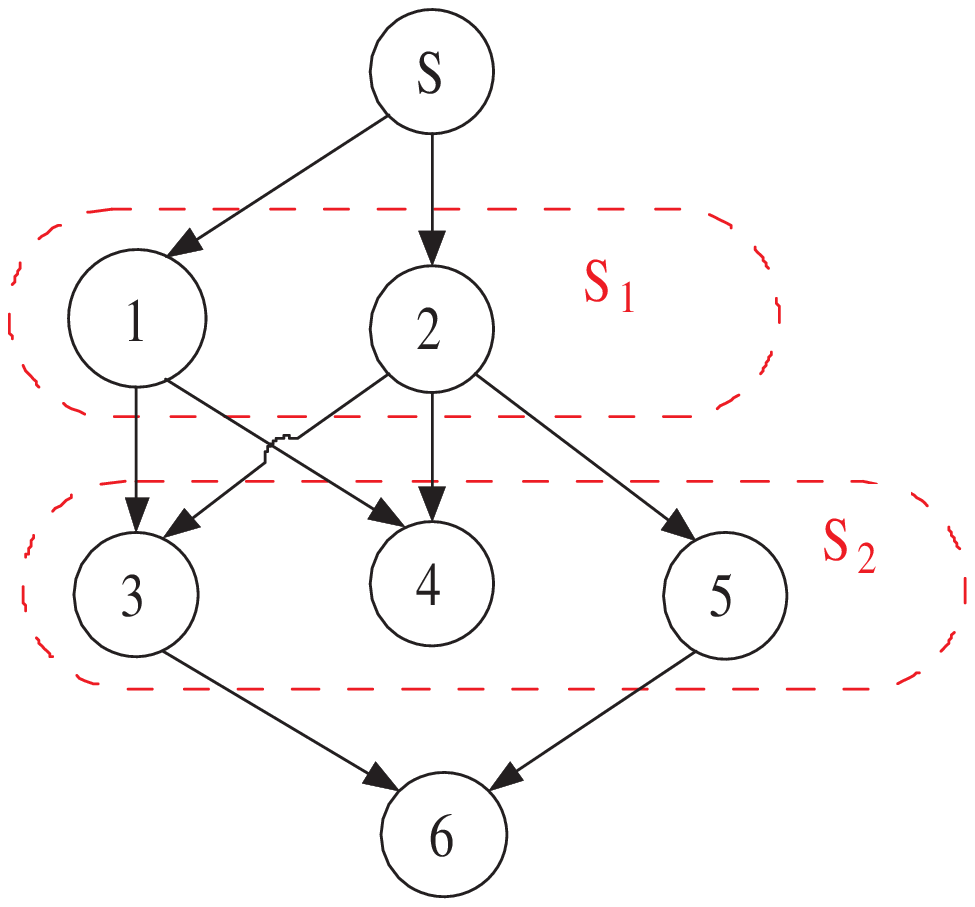}
  \caption{A strictly hierarchical P2P network}\label{fig:SHP2P}
\end{figure}

Fig.~\ref{fig:SHP2P} shows a strictly hierarchical P2P network with
3 levels. In a strictly hierarchical P2P network, peers in level $K$
work together as a virtual server and upload to peers in level
$K+1$. In Fig.~\ref{fig:SHP2P}, peer 1 and 2 form the virtual server
in level $1$, denoted as $S_1$. Peer 3, 4 and 5 form the virtual
server in level $2$, denoted as $S_2$. Since all transmission flows
are between two consecutive levels, peer selection algorithms and
chunk selection algorithms can be locally designed level by level.

\begin{figure}
  \centering
  \includegraphics[width=0.35\textwidth]{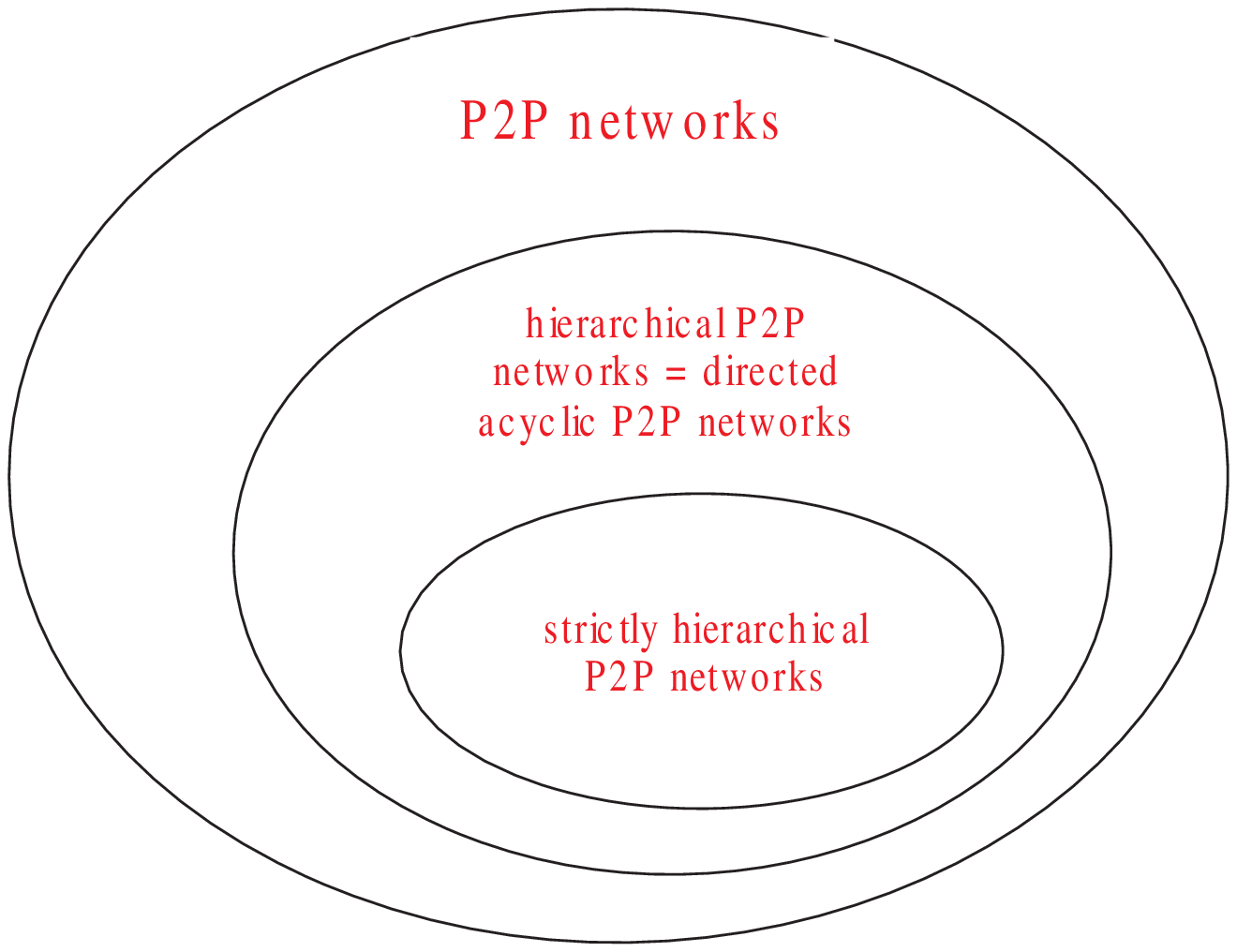}
  \caption{Conclusion on taxonomy of overlay networks}\label{fig:relation}
\end{figure}

The relationships among the P2P network, the hierarchical P2P
network and the strictly hierarchical P2P network are concluded in
Fig.~\ref{fig:relation}.

\subsection{Network of sub-peers}
\label{sec:subpeer}

\begin{figure}
  \centering
  \includegraphics[width=0.35\textwidth]{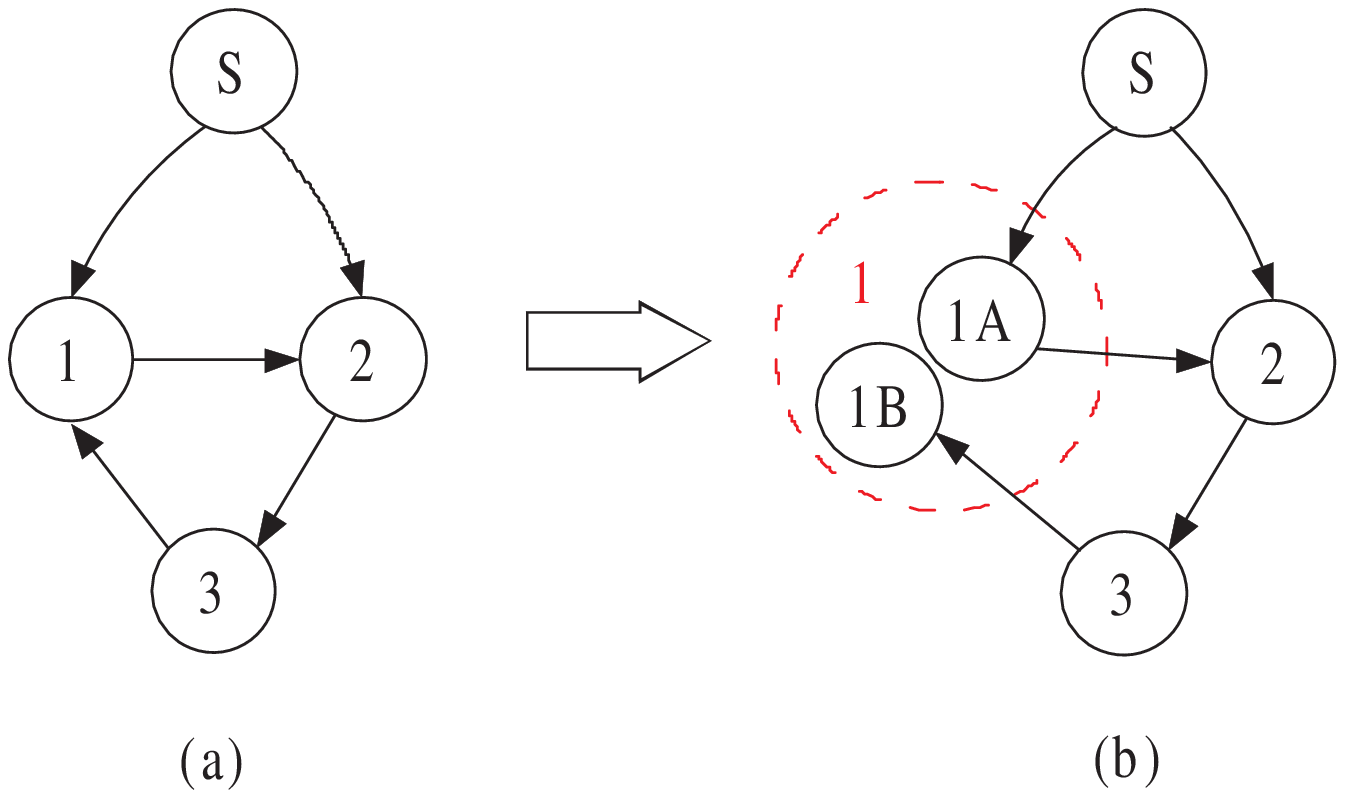}
  \caption{Peer division}\label{fig:peerdivision}
\end{figure}

A P2P overlay network is divisible if peers can be divided into
sub-peers, and it is indivisible if peers cannot be divided. The
division of peers can be performed at the application layer
\cite{AND04} \cite{S.WEEbook}. Even for indivisible P2P network,
peers can be conceptually divided into virtual sub-peers for
theoretical analysis. A simple example of peer division is given in
Fig.~\ref{fig:peerdivision}. Fig.~\ref{fig:peerdivision}(a) shows
the original P2P network and Fig.~\ref{fig:peerdivision}(b) shows
the network of sub-peers after the division of peer 1 into sub-peer
$1A$ and sub-peer $1B$.

A peer division is equivalent if the transmission rates from the
server to each peer and from each peer to each other peer are
invariant. The network of sub-peers is equivalent to the original
P2P network if the peer division is equivalent. The peer division in
Fig.~\ref{fig:peerdivision} is equivalent if $r_{s \rightarrow 1} =
r_{s \rightarrow 1A}$, $r_{1 \rightarrow 2} = r_{1A \rightarrow 2}$
and $r_{3 \rightarrow 1} = r_{3 \rightarrow 1B}$.

\newtheorem{main}[HP2P=NoCycle]{Theorem}
\begin{main}\label{theorem:main}
Any P2P network with $N$ peers and $K$ levels can be decomposed into
an \emph{equivalent} network of sub-peers that is strictly
hierarchical, and has at most $K$ levels, each of which contains at
most $N$ sub-peers.
\end{main}

\begin{figure}
  \centering
  \includegraphics[width=0.45\textwidth]{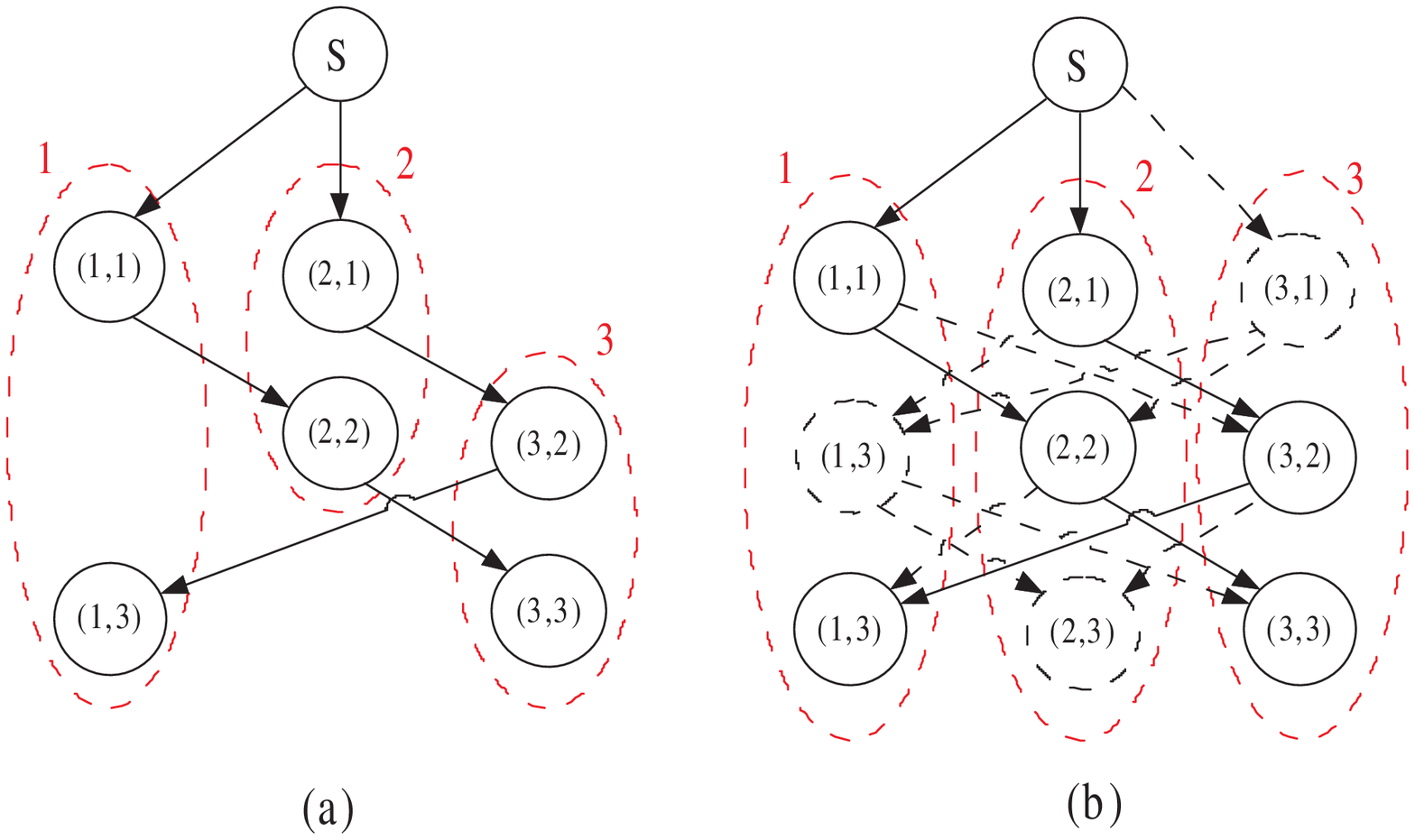}
  \caption{Construction of the network of sub-peers}\label{fig:complexfulldivision}
\end{figure}

\emph{Proof of Theorem \ref{theorem:main}:} In order to prove the
theorem, it suffices to construct a strictly hierarchical network of
sub-peers which is equivalent to the original P2P network. For any
P2P network with $N$ peers and $K$ levels, denote the server $S$ as
node $0$ and the peers as node $1, 2, \cdots , N$. Let
\begin{align}
K_i = & \{k : \exists \textrm{ a directed acyclic path from } S
\nonumber \\
& \textrm{ to peer } i \textrm{ with length } k\},
\end{align}
and so $|K_i| \leq K$ is the cardinality of $K_i$. Divide peer $i$
into $|K_i|$ sub-peers, which are denoted as sub-peer $(i,k)$ for $k
\in K_i$. The sub-peer $(i,k)$ is the part of peer $i$ in level $k$.
In the network of sub-peers, there is an edge from the server $S$ to
sub-peer $(i,1)$ if and only if $S \rightarrow i$ in the original
network. We assign the transmission rate $r_{s \rightarrow (i,1)} =
r_{s \rightarrow i}$. There is an edge from sub-peer $(i,k)$ to
sub-peer $(j,k+1)$ if and only if there exits a directed acyclic
path with length $k+1$ such that $S \rightarrow \cdots \rightarrow i
\rightarrow j$ in the original P2P network. We assign the
transmission rates level by level with
\begin{align}
r_{(i,k)} & = \sum_{(i',k-1) \rightarrow (i,k)} r_{(i',k-1)
\rightarrow (i,k)},  \label{eq:assign1} \\
r_{i \rightarrow j}^{k} &
= r_{i \rightarrow j} - \sum_{m \leq k} r_{(i,m-1) \rightarrow (j,m)}, \label{eq:assign2}\\
r_{(i,k) \rightarrow (j,k+1)} & = \min ( r_{(i,k)} - \sum_{j' < j}
r_{(i,k) \rightarrow (j',k+1)}, r_{i \rightarrow j}^{k}),
\label{eq:assign3}
\end{align}
where $r_{(i,k)}$ is the total download rate of the sub-peer
$(i,k)$, $r_{i \rightarrow j}$ is the transmission rate from peer
$i$ to peer $j$ in the original P2P network, $r_{i \rightarrow
j}^{k}$ is the unassigned transmission rate from peer $i$ to peer
$j$ until level $k$ in the network of sub-peers, and $r_{(i,k)
\rightarrow (j,k+1)}$ is the transmission rate from $(i,k)$ to
$(j,k+1)$. This division for the network in
Fig.\ref{fig:peerdivision}(a) is shown in
Fig.~\ref{fig:complexfulldivision}(a).

Since $|K_i| \leq K$ for all $i$, the constructed network of
sub-peers has at most $K$ levels. There are at most $N$ sub-peers in
each level because each level contains at most one sub-peer of any
peer. It is also easy to check that the constructed network of
sub-peers is strictly hierarchical.

In order to prove that the network of sub-peers is equivalent to the
original P2P network, we need to show that
\begin{equation}
r_{i \rightarrow j} = \sum_{k \leq K} r_{(i,k-1) \rightarrow (j,k)}.
\end{equation}
If $i = 0$, i.e., node $i$ the server $S$, then
\begin{equation}
r_{S \rightarrow j} = r_{S \rightarrow (j,1)} = \sum_{k \leq K}
r_{(0,k-1) \rightarrow (j,k)}.
\end{equation}
If $i \neq 0$, suppose for all $k \leq K$,
\begin{align}
r_{(i,k-1) \rightarrow (j,k)} & = \min ( r_{(i,k-1)} - \sum_{j' < j}
r_{(i,k-1) \rightarrow (j',k)}, r_{i \rightarrow j}^{k-1}) \nonumber
\\
& < r_{i \rightarrow j}^{k-1},
\end{align}
then
\begin{equation}
\sum_{k = 2}^{K} (r_{(i,k-1)} - \sum_{j' < j} r_{(i,k-1) \rightarrow
(j',k)}) < \sum_{k = 2}^{K-1} r_{(i,k-1) \rightarrow (j,k)} + r_{i
\rightarrow j}^{K-1},
\end{equation}
and so
\begin{equation}
r_{i} < \sum_{j' < j} r_{i \rightarrow j'} + r_{i \rightarrow j} =
\sum_{j' \leq j} r_{i \rightarrow j'},
\end{equation}
which is infeasible. Hence, there always exists some $m \leq K$ such
that
\begin{align}
r_{(i,m-1) \rightarrow (j,m)} & = \min ( r_{(i,m-1)} - \sum_{j' < j}
r_{(i,m-1) \rightarrow (j',m)}, r_{i \rightarrow j}^{m-1}) \nonumber
\\
& = r_{i \rightarrow j}^{m-1},
\end{align}
then $r_{i \rightarrow j}^{k} = 0$ for $ k \geq m$, $r_{(i,k-1)
\rightarrow (j,k)} = 0$ for $k > m$, and hence
\begin{align}
\sum_{k \leq K} r_{(i,k-1) \rightarrow (j,k)} & = \sum_{k \leq m}
r_{(i,k-1) \rightarrow (j,k)} \nonumber \\
& = r_{i \rightarrow j}.
\end{align}

Finally, we need to show that the assigned transmission rates in the
network of sub-peers are feasible, i.e., the download rate of each
sub-peer is greater than or equal to it upload rate. Plugging $j=N$
into (\ref{eq:assign3}),
\begin{equation}
r_{(i,k) \rightarrow (N,k+1)} \leq  r_{(i,k)} - \sum_{j' < N}
r_{(i,k) \rightarrow (j',k+1)},
\end{equation}
and so
\begin{equation}
r_{(i,k)} \geq \sum_{j' \leq N} r_{(i,k) \rightarrow (j',k+1)}.
\end{equation}
Q.E.D.

If edges with zero transmission rate and sub-peers with zero
download bandwidth and zero upload bandwidth are allowed, divide
peer $i$ into $K$ sub-peers, which are denoted as sub-peer $(i,k)$
for $k = 1,\cdots,K$. The sub-peer $(i,k)$ is the part of peer $i$
in level $k$. There is an edge from the server $S$ to sub-peer
$(i,1)$ for all $i$. There is an edge from sub-peer $(i,k)$ to
sub-peer $(j,k+1)$ for all $i \neq j$. The network of sub-peers
after this division is also strictly hierarchical, and keeps all the
transmission paths in the original network. This peer division for
the P2P network in Fig.~\ref{fig:peerdivision}(a) is shown in
Fig.~\ref{fig:complexfulldivision}(b), in which the dotted border
sub-peers have zero download bandwidth and zero upload bandwidth,
and the dotted edges have zero transmission rate. By Theorem
\ref{theorem:main}, there exits a transmission rate assignment such
that the network of sub-peers after this peer division is equivalent
to the original P2P network. Hence, minimizing the WADT for strictly
hierarchical networks of sub-peers is equivalent to minimizing the
WADT for P2P networks.

In a very large P2P network with $N$ peers, the length of the
longest directed acyclic path $K = \max_i |K_i|$ is usually much
less than $N$. Denote $d_{i,j}, u_{i,j}, r_{i,j}$ as the download
bandwidth, upload bandwidth and the download rate of the part of
peer $i$ in level $j$, where $d_{i,j} \geq u_{i,j}$ and $d_{i,j}
\geq r_{i,j}$. The sub-peers in level $j$ work together as a virtual
server to upload to the sub-peers in level $j+1$. Denote $S_j$ as
the upload bandwidth of the virtual server in level $j$, $1 \leq j
\leq K$, and so $S_j = \sum_{i=1}^{N} \min(u_{i,j},r_{i,j})$. Let
$S_0 = S$.

\newtheorem{complexcompactrate}[P2P]{Lemma}
\begin{complexcompactrate}\label{theorem:complexcompactrate}
Given the server with bandwidth $S$ and the peers with download
bandwidth $d_i$ and upload bandwidth $u_i$, $1 \leq i \leq N$, there
exists an optimal strictly hierarchical network of sub-peers which
achieves the minimum WADT such that $r_{i,j} \geq u_{i,j}$ for all
$1 \leq i \leq N$, $1 \leq j \leq K-1$.
\end{complexcompactrate}

\emph{Proof:} Suppose $d_{i,j}, u_{i,j}, r_{i,j}$ determine an
optimal strictly hierarchical network of sub-peers which connects
the server and the peers, and achieves the minimum WADT. It is clear
that $\tilde{d}_{i,j}, \tilde{u}_{i,j}, \tilde{r}_{i,j}$ is also an
optimal strictly hierarchical network of sub-peers which has
$\tilde{r}_{i,j} \geq \tilde{u}_{i,j}$ for any $1 \leq i \leq N$, $1
\leq j \leq K-1$, where
\begin{equation}
\tilde{r}_{i,j}=r_{i,j}
\end{equation}
\begin{equation}
\tilde{u}_{i,j} = \left\{ \begin{array}{ll}
\min(u_{i,j}, r_{i,j}) & \textrm{for $1 \leq j \leq K-1$}\\
u_i-\sum_{j=1}^{K-1}\tilde{u}_{i,j} & \textrm{for $j=K$}
\end{array} \right.
\end{equation}
\begin{equation}
\tilde{d}_{i,j} = \left\{ \begin{array}{ll}
r_{i,j} & \textrm{for $1 \leq j \leq K-1$}\\
d_i-\sum_{j=1}^{K-1}\tilde{d}_{i,j} & \textrm{for $j=K$}
\end{array} \right.
\end{equation}
The theorem is immediately proved. Q.E.D.

\newtheorem{complexcompactserver}{Corollary}
\begin{complexcompactserver}\label{theorem:complexcompactserver}
$S_j = \sum_{i=i}^N \min(u_{i,j},r_{i,j}) = \sum_{i=i}^N u_{i,j}$
for any $1 \leq j \leq K-1$.
\end{complexcompactserver}

\section{Optimizing and Bounding Weighted Average Download Time}
\label{sec:WADT}

A challenging problem for content distribution applications is how
to build a P2P content distribution network to achieve the minimum
weighted average download time. The WADT of a P2P network with $N$
peers is $\sum_{i=1}^{N}w_i/r_i$, where $w_i$ is the weight and
$r_i$ is the download rate for peer $i$. The weights $w_i$,
$i=1,\cdots,N$, depend on the application. For broadcast
applications such as CoolStreaming \cite{Zhang05} and Overcast
\cite{Overcast}, in which all peers in the P2P network are
interested in the same content, all weights of peers in the content
distribution system can be set to be 1. For multicast applications
such as ``Tribler'' \cite{Tribler}, in which some helpers, who are
not interested in any particular content, store part of the content
and share with other peers, weights of helpers are 0 and weights of
other peers are 1. In some applications, P2P systems can also
partition peers into several classes and assign different weights to
peers in different classes.

\subsection{Building Optimal P2P Networks Using Convex Programming}

Theorem \ref{theorem:main} demonstrates that minimizing the WADT for
P2P networks with $N$ peers and $K$ levels is equivalent to
minimizing the WADT for strictly hierarchical networks of sub-peers
with $K$ levels, each of which contains $N$ sub-peers. $K \ll N$ for
large P2P networks. Recall that the bandwidth of the server is
$S=S_0$. The download bandwidth and upload bandwidth of peer $i$ are
$d_i$ and $u_i$ respectively, where $d_i \geq u_i$. $S_j$ is the
bandwidth of the virtual server in level $j$, $1 \leq j \leq K$.
$d_{i,j}, u_{i,j}, r_{i,j}$ are the download bandwidth, the upload
bandwidth and the download rate of the part of peer $i$ in level
$j$.

The constraints on the download bandwidths and upload bandwidths are
$d_{i,j} \geq u_{i,j}$ for any $1 \leq i \leq N$ and $1 \leq j \leq
K$, $\sum_{j=1}^K d_{i,j}=d_i$ and $\sum_{j=1}^K u_{i,j}=u_i$ for
any $1 \leq i \leq N$. The constraints on the download rates are
$r_{i,j} \leq d_{i,j}$ for any $1 \leq i \leq N$ and $1 \leq j \leq
K$. Lemma \ref{theorem:complexcompactrate} shows that any optimal
strictly hierarchical network of sub-peers has $r_{i,j} \geq
u_{i,j}$ for any $1 \leq i \leq N$ and $1 \leq j \leq K-1$.
$\sum_{j=1}^K r_{i,j}=r_i$ is the total download rate for peer $i$.
Corollary \ref{theorem:complexcompactserver} demonstrates that the
bandwidth of the virtual server in level $j$ is $S_j=\sum_{i=1}^N
u_{i,j}$ for $j=1,\cdots,K-1$. One has $S_{j-1} \geq \sum_{i=1}^N
r_{i,j}$ since the virtual server in Level $j-1$ uploads content to
the sub-peers in Level $j$.

All the constraints are linear and the objective function,
$\sum_{i=1}^N w_i / r_i$, is a convex function. Therefore, the
problem of minimizing the WADT is a convex optimization program,
which can be solved by convex optimization solvers such as the
interior-point method. The complexity of the interior-point method
to solve this problem is $o(\sqrt{K^3N^3})$ \cite{BookConvexOpt}.
The solution to the convex optimization program provides the
topology of the optimal P2P network which achieves the minimum WADT,
however, the number of the peers is usually very large for most P2P
applications and so the complexity for solving the convex program is
too high. Fortunately, this convex program has an analytical
achievable upper bound which is very tight to the optimal solution
and its complexity is linear to the number of the peers in a P2P
network.

\subsection{Analytical Upper Bound and Lower Bound for the Convex Optimization Program}
\begin{figure}
  \centering
  \includegraphics[width=0.15\textwidth]{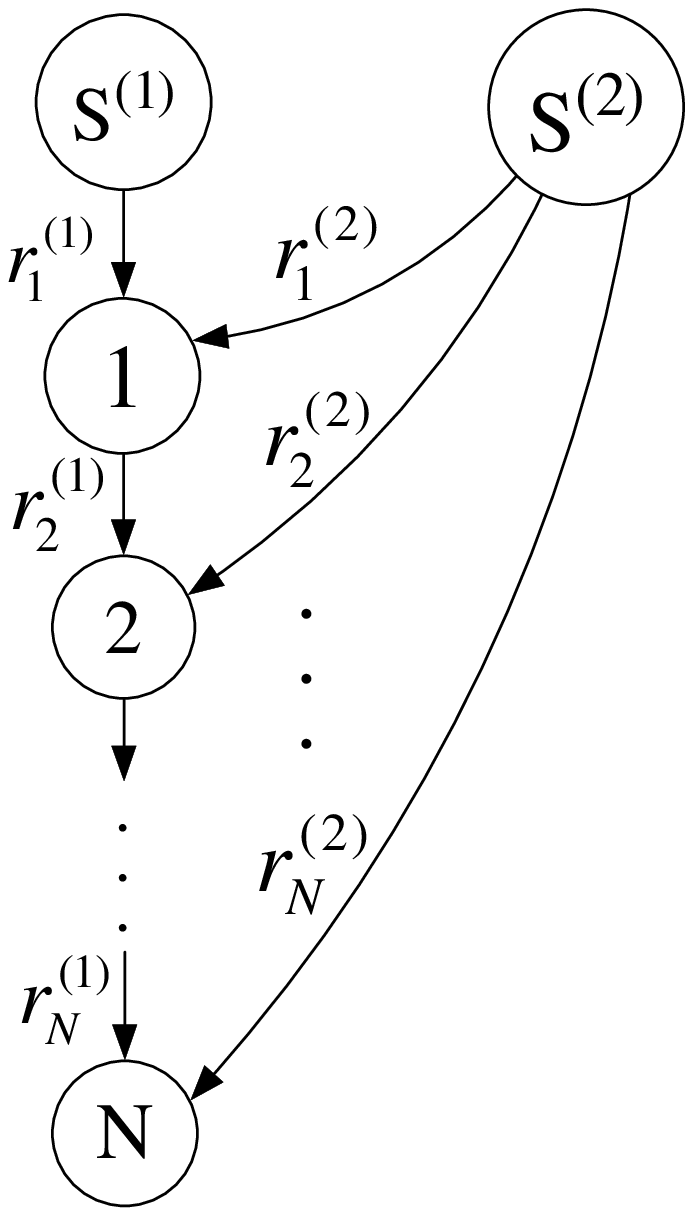}
  \caption{A hierarchical P2P network providing the upper bound}\label{fig:upperbound}
\end{figure}

Theorem \ref{theorem:HP2P=NoCycle} shows that any hierarchical P2P
network contains no directed cycle and vice versa. Constructing a
hierarchical P2P network shown in Fig.\ref{fig:upperbound} provides
an achievable upper bound for the minimum WADT. Assume that $S \gg
\max_{i}(u_i)$, which is always true practically. Without loss of
generality, assume $u_1 \geq u_2 \geq \cdots \geq u_N$. Partition
the server into two parts $S^{(1)}=\max_{i}(u_i) = u_1$ and $S^{(2)}
= S -\max_{i}(u_i) = S - u_1$. The first part $S^{(1)}$ can serve
all the peers with $r_i^{(1)} = u_i$. The second part $S^{(2)}$ can
then freely serve for the peers only with the constraints $d_i \geq
r_i^{(1)}+r_i^{(2)} \geq u_i$ and $S^{(2)} \geq \sum_{i}^N
r_i^{(2)}$. Thus, the upper bound is the solution to the
optimization problem
\begin{align}
\textrm{Minimize} & \quad \sum_{i=1}^{N} w_i / r_i \label{eq:upperbound}\\
\textrm{Subject to} & \quad d_i \geq r_i \geq u_i, i=1,2,\cdots, N, \nonumber \\
&\quad \sum_{i=1}^{N} r_i -u_i \leq S - \max_i(u_i). \nonumber
\end{align}

The Karush-Kuhn-Tucker conditions for the optimal solution are
\begin{align}
 &\mu_i, \nu_i, \lambda \geq 0, \nonumber \\
 &-w_i r_i^{-2}+\mu_i-\nu_i+\lambda = 0, \nonumber\\
 &\mu_i (r_i-d_i)= 0, \nonumber \\
 &\nu_i (u_i-r_i)= 0, \nonumber \\
 &\lambda \left(\sum_{i=1}^{N} (r_i-u_i) - (S- \max_i(u_i)) \right ) = 0. \nonumber
\end{align}

Combining the the Karush-Kuhn-Tucker conditions and the constraints
in the problem, we obtain the upper bound which is $\sum_{i=1}^{N}
w_i / r_i$ with
\begin{equation}\label{eq:optsimplefull}
r_i = \left\{ \begin{array}{ll}
\sqrt{w_i}R & \textrm{if $u_i \leq \sqrt{w_i}R \leq d_i$}\\
u_i & \textrm{if $u_i > \sqrt{w_i}R$}\\
d_i & \textrm{if $\sqrt{w_i}R > d_i$}
\end{array} \right.
\end{equation}
where $R$ is chosen appropriately such that $\sum_{i=1}^{N}
(r_i-u_i) = S - \max_{i}(u_i)$.

This upper bound is almost optimal. Consider the following lower
bound which is very close to the upper bound. Since $S \geq
\sum_{i=1}^{N}(r_i - u_i)$ is a relaxed constraint, an lower bound
is the solution to the optimization problem
\begin{align}
\textrm{Minimize} & \quad \sum_{i=1}^{N} w_i / r_i \label{eq:lowerbound}\\
\textrm{Subject to} & \quad d_i \geq r_i \geq u_i, i=1,2,\cdots, N, \nonumber \\
&\quad \sum_{i=1}^{N} r_i -u_i \leq S. \nonumber
\end{align}

Note that this optimization problem (\ref{eq:lowerbound}) differs
from the problem (\ref{eq:upperbound}) only by replacing $S$ with $
S - \max_{i}(u_i)$. Thus, the lower bound is $\sum_{i=1}^{N} w_i /
r_i$ with
\begin{equation}
r_i = \left\{ \begin{array}{ll}
\sqrt{w_i}R & \textrm{if $u_i \leq \sqrt{w_i}R \leq d_i$}\\
u_i & \textrm{if $u_i > \sqrt{w_i}R$}\\
d_i & \textrm{if $\sqrt{w_i}R > d_i$}
\end{array} \right.
\end{equation}
where $R$ is chosen appropriately such that $\sum_{i=1}^{N}
(r_i-u_i) = S$.

Since $S \gg \max(u_i)$, one has $S \simeq S-\max_{i}(u_i)$. The
upper bound and the lower bound are almost the same. Therefore, both
of these two bounds are very tight, and the upper bound is almost an
analytical solution to minimizing the WADT. Hence, it is sufficient
to build a hierarchical P2P network to achieve the minimum WADT.

\subsection{Practical Solutions}

Compared with building networks of sub-peers and hierarchical P2P
networks, it is more practical to build a strictly hierarchical P2P
network. First, in a strictly hierarchical P2P network, peers do not
need to be decomposed into sub-peers. Thus, it can have a much
simpler protocol in network layer than a network of sub-peers.
Second, in a strictly hierarchical P2P network, each level usually
contains only tens or hundreds of peers. Therefore, one can locally
design peer selection algorithms and chunk selection algorithms
level by level, which only depend on a small collection of peers in
the P2P network. The locally designed peer and chunk selection
algorithm might be much simpler than the network coding strategy and
other global designed peer and chunk selection algorithms. The
following theorem shows that there exists a strictly hierarchical
network of sub-peers which achieves the upper bound in the previous
subsection, and is very close to a strictly hierarchical P2P
network.

\newtheorem{almostoptimal}[HP2P=NoCycle]{Theorem}
\begin{almostoptimal}\label{theorem:almostoptimal}
There exist multiple upper-bound-achieving networks of sub-peers in
which at most $K-1$ peers need to be decomposed into 2 sub-peers,
where $K \ll N$ is the number of levels.
\end{almostoptimal}

\emph{Proof}: Equation (\ref{eq:optsimplefull}) gives $r_i$,
$i=1,\cdots,N$, for the hierarchical P2P network which achieves the
upper bound. Convert the hierarchical P2P network to multiple
optimal strictly hierarchical networks of sub-peers by the following
algorithm.

\begin{algorithm}[h]
\caption{Peer Placement Algorithm} \label{alg:peer_placement}
\begin{algorithmic} [1]
    \STATE Initialize level $l=1$;
    \STATE Initialize the rest server bandwidth for the current level
    $s=S$;
    \STATE Initialize the set of the rest of the peers $G={1,2,\cdots,N}$;
    \WHILE {$G$ is not empty}
    \STATE Choose any peer $i$ from the set $G$;
    \IF{$s \geq r_i$}
        \STATE Put Peer $i$ in the current level $l$;
        \STATE Set $G = G\backslash \{i \}$;
        \STATE Set $s = s - r_i$;
    \ELSE
        \STATE Decompose Peer $i$ into 2 sub-peers with $r_i^{(1)} = s$, $r_i^{(2)} = r_i - s$, $u_i^{(1)} = \min(u_i,r_i^{(1)})$, $u_i^{(2)} = u_i -
        \min(u_i,r_i^{(1)})$;
        \STATE Put the sub-peer $(1)$ in level $l$ and the sub-peer $(2)$ in level
        $l+1$;
        \STATE Set $s = (\sum _{\textrm{Peer } j \textrm{ in level }l} u_j)
- r_i^{(2)}$;
        \STATE Set $l = l+1$;
        \STATE Set $G = G\backslash \{i \}$;
    \ENDIF
    \ENDWHILE
\end{algorithmic}
\end{algorithm}

This algorithm indicates that at most $K-1$ peers need to be divided
into 2 sub-peers and each level of the constructed network of
sub-peers contains at most 2 sub-peers. Since there are multiple
choices in step (2), this algorithm provides multiple strictly
hierarchical networks of sub-peers which are very close to a
strictly hierarchical P2P network. Q.E.D.

Another practical solution is to solve the problem of minimizing the
WADT directly for strictly hierarchical P2P networks. This problem
can be formulated as a 0-1 convex optimization. The complexity to
solve a 0-1 convex optimization problem is exponential to the
problem size.

Fortunately, this problem has an analytical suboptimal solution.
Suppose there are $N$ peers and $K$ levels in a strictly
hierarchical P2P network and the level location of each peer is
already given. Then the global optimization problem can be
decomposed into $K$ local optimization problem and all of them have
an analytical ``Water Filling'' solution.

In particular, suppose the bandwidth of the server is $S=S_0$. The
download bandwidth, upload bandwidth and the allocated download rate
of Peer $i$ are $d_i$, $u_i$ and $r_i$ respectively. $S_j$, the
bandwidth of the virtual server in Level $j$, is equal to the
summation of the upload bandwidths of the peers in Level $j$. If
$S_j$ is greater or equal to the summation of the upload bandwidth
of all peers in level $j+1$, then one has $d_i \geq r_i \geq u_i$
and $S_{j+1} = \sum _{\textrm{Peer } i \textrm{ in level }j} u_i$.
If $S_j$ is less than the summation of the upload bandwidth of all
peers in level $j+1$, one has $0 \leq r_i \leq u_i$ and $S_{j+1} =
S_{j} = \sum _{\textrm{Peer } i \textrm{ in level }j} r_i$.
Therefore, no matter how to allocate download rates to peers in each
level, the bandwidth of the virtual server in each level is fixed as
long as the level positions of peers are fixed. Thus, the global
minimum weighted average transmission rate problem can be decomposed
to $K$ local optimization problem. Each of them has a fixed server
bandwidth and fixed number of peers in the level. Thus, if the
virtual server $S_j$ is greater or equal to the summation of the
upload bandwidth of all peers in level $j+1$, then the local
optimization problem is
\begin{align}\label{eq:optimization}
\textrm{Minimize} & \quad \sum_{i=1}^{N_{j+1}} w_i / r_i \nonumber\\
\textrm{Subject to} & \quad d_i \geq r_i \geq u_i, i=1,2,\cdots, N_{j+1} \nonumber \\
&\quad \sum_{i=1}^{N_{j+1}} r_i \leq S_j \nonumber
\end{align}
where $N_{j+1}$ is the number of peers in level $j+1$, for
$j=0,\cdots,K-1$.

By the Karush-Kuhn-Tucker conditions, the optimal solution is
\begin{equation}
r_i = \left\{ \begin{array}{ll}
\sqrt{w_i}R & \textrm{if $u_i \leq \sqrt{w_i}R \leq d_i$}\\
u_i & \textrm{if $u_i > \sqrt{w_i}R$}\\
d_i & \textrm{if $\sqrt{w_i}R > d_i$}
\end{array} \right.
\end{equation}
where $R$ is chosen appropriately such that $\sum_{i=1}^{N_{j+1}}
r_i = S_j$.

If the virtual server $S_j$ for level $j+1$ is less than the
summation of the upload bandwidth of all peers in level $j+1$, then
the local optimization problem is
\begin{align}\label{eq:optimization}
\textrm{Minimize} & \quad \sum_{i=1}^{N_{j+1}} w_i / r_i \nonumber\\
\textrm{Subject to} & \quad u_i \geq r_i \geq 0, i=1,2,\cdots, N_{j+1} \nonumber \\
&\quad \sum_{i=1}^{N_{j+1}} r_i \leq S_j \nonumber
\end{align}
where $N_{j+1}$ is the number of peers in level $j+1$, for
$j=0,\cdots,K-1$.

By the Karush-Kuhn-Tucker conditions, the optimal solution is
\begin{equation}
r_i = \left\{ \begin{array}{ll}
\sqrt{w_i}R & \textrm{if $0 \leq \sqrt{w_i}R \leq u_i$}\\
u_i & \textrm{if $u_i < \sqrt{w_i}R$}
\end{array} \right.
\end{equation}
where $R$ is chosen appropriately such that $\sum_{i=1}^{N_{j+1}}
r_i = S_j$.

\section{Results}
\label{sec:results}

In this section, we simulate and evaluate the performances of the
methods provided in Section \ref{sec:WADT}. Note that in this paper
the download bandwidth $d$ and the upload bandwidth $u$  are decided
at the application layer instead of the physical layer. Thus, the
download bandwidth $d$ can be continuously distributed in a large
range such as 10 kbps to 100 Mbps. We assume that the download
bandwidths $d_i$, $i=1,\cdots,N$ are independently identically
distributed (i.i.d.) with uniform distribution over
$[\beta,2-\beta]$, where $\beta$ is a very small positive number
such as $10^{-2},10^{-4}$. This distribution has the normalized mean
value 1 and large maximum to minimum ratio $(2-\beta)/\beta$. In
practice, this parameter $\beta$ can be determined as the minimum
download bandwidth which is required by the P2P system. We also
assume that the upload bandwidth $u_i$ is uniformly distributed over
$[\alpha d_i,d_i]$, where $0<\alpha<1$ is the minimum upload to
download bandwidth required by the P2P system.

\subsection{A small network simulation}
In this small simulation, the bandwidth of the server is 10 and the
number of the peers in the network is $N=100$. The download
bandwidths $d_i$, $i=1,\cdots,100$, are i.i.d. with uniform
distribution over $[0.01,1.99]$, i.e., $\beta = 10^{-2}$. For each
peer $i$, the upload bandwidth is uniformly distributed over
$[0.1d_i,d_i]$, i.e., $\alpha = 0.1$. The weights for all the peers
are equal and normalized to be $1/N = 1/100$, and so $\sum_i w_i =
1$. Thus, the WADT is $\sum_{i} w_i/r_i$. We will compare 7 methods
in this experiment. They are
\begin{itemize}
\item 1. Optimal solution to the convex optimization program with level $K=30$. The convex program solver is
CVX \cite{CVX}.
\item 2. The upper bound (\ref{eq:upperbound}) by constructing a hierarchical P2P network.
\item 3. The lower bound (\ref{eq:lowerbound}) by relaxing the
constraints.
\item 4. The suboptimal solution to the 0-1 convex optimization
problem for strictly hierarchical P2P networks. The number of the
levels are set to be $K=30$ and we randomly put around $N/K$ peers
in each level.
\item 5. The suboptimal solution to the 0-1 convex optimization
problem for strictly hierarchical P2P networks. We convert the
upper-bound-achieving hierarchical P2P network (Method 2) to a
network of sub-peers by Algorithm \ref{alg:peer_placement}. The
peers are placed into levels according to the constructed network of
sub-peers, which is very close to a strictly hierarchical P2P
network.
\item 6 This is a trivial upper bound such that the rate of each
peer is the same as the upload bandwidth.
\item 7 This is a trivial lower bound such that the rate of each
peer is the same as the download bandwidth.
\end{itemize}

\begin{figure}
  \centering
  \includegraphics[width=0.5\textwidth]{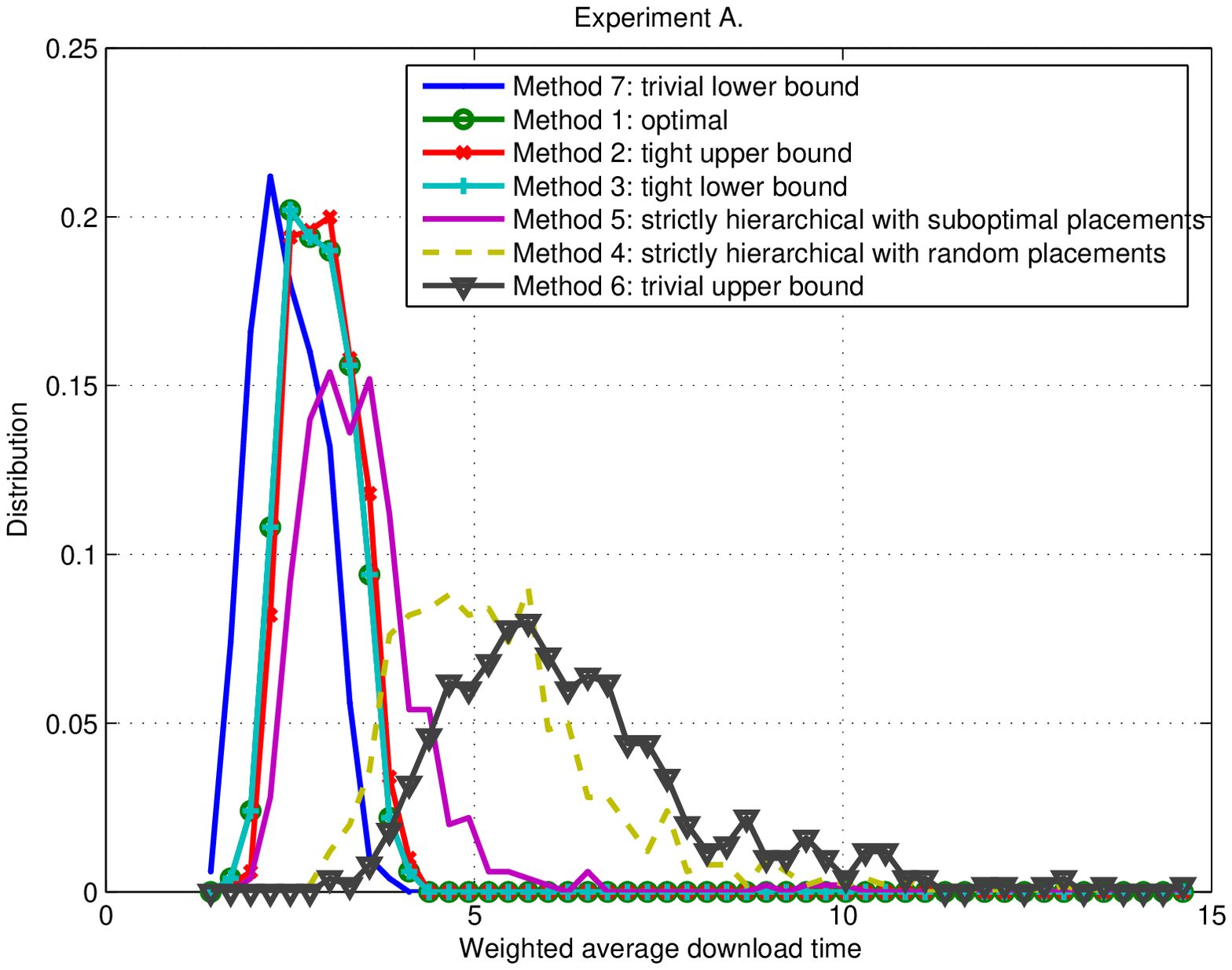}
  \caption{Distribution of the WADT for different
  methods.}\label{fig:TimeA}
\end{figure}

\begin{figure}
  \centering
  \includegraphics[width=0.5\textwidth]{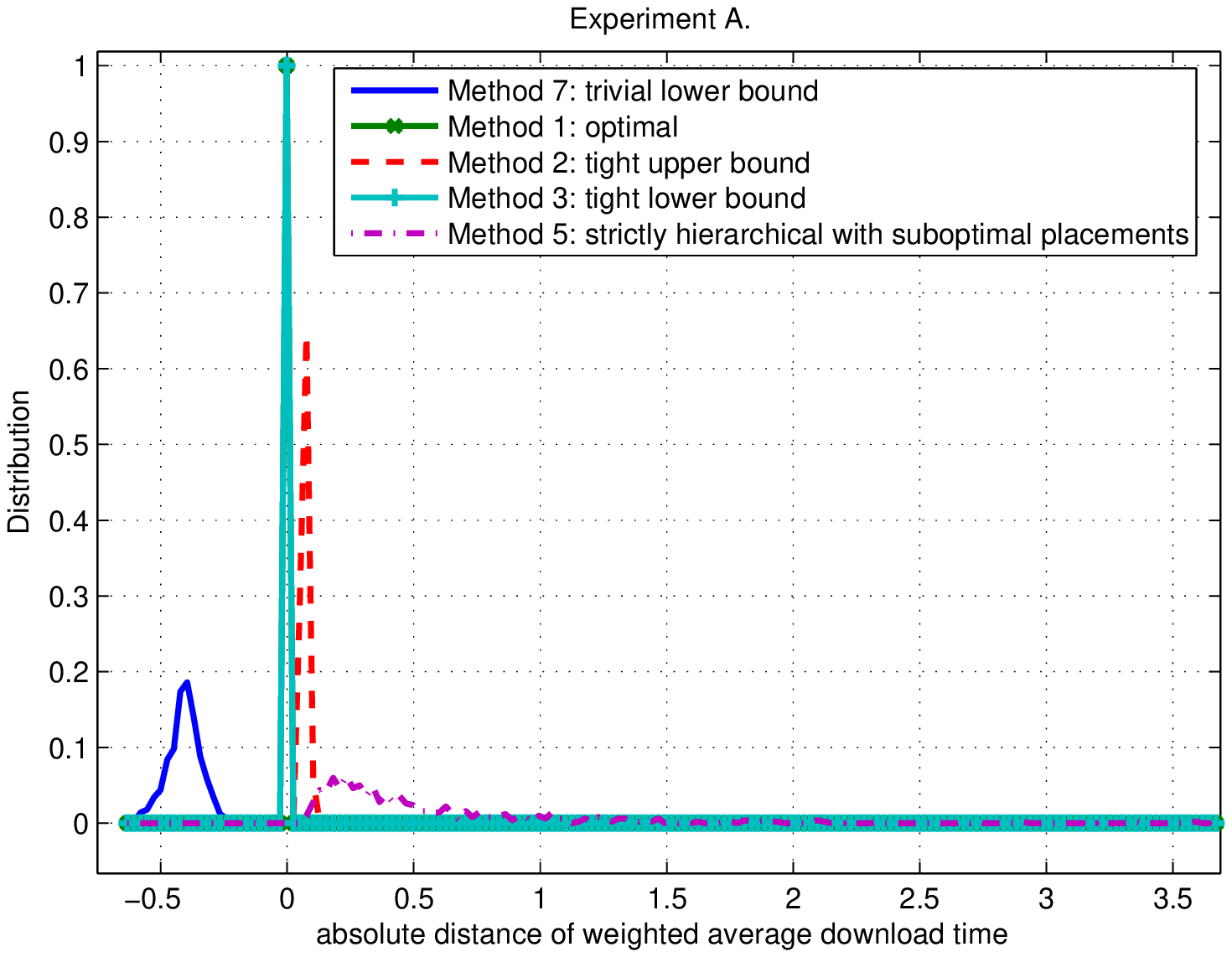}
  \caption{Distribution of the differences of the WADT between Method 3 and other methods.}\label{fig:TimeDiffA}
\end{figure}

\begin{figure}
  \centering
  \includegraphics[width=0.5\textwidth]{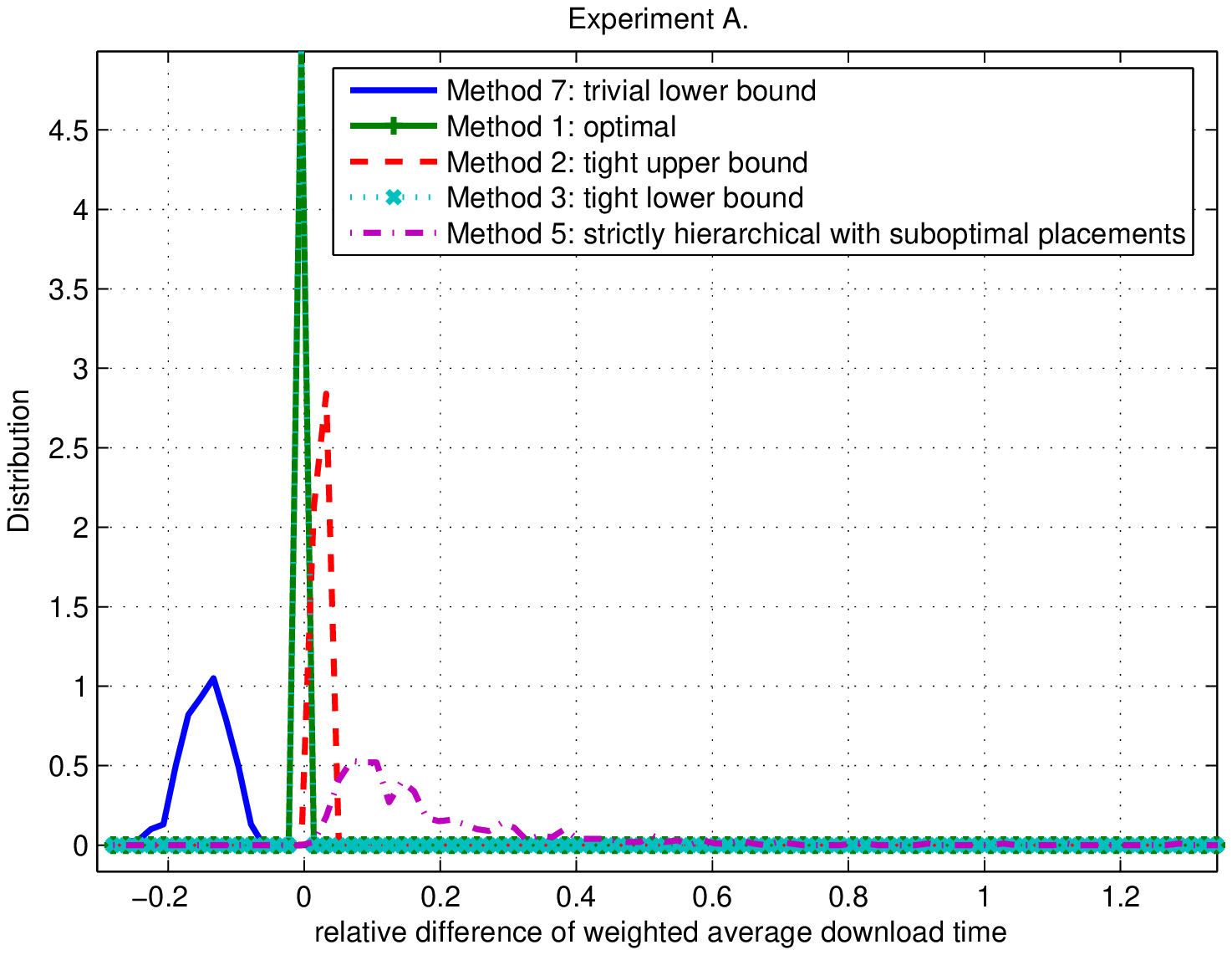}
  \caption{Distribution of the relative difference of the WADT between Method 3 and other methods.}\label{fig:RelTimeDiffA}
\end{figure}

The distribution of the weighted average download times of 500
experiments are shown in Fig.~\ref{fig:TimeA}. The distribution of
the difference and the relative difference between Method 3 and
other methods are shown in Fig.~\ref{fig:TimeDiffA} and
Fig.~\ref{fig:RelTimeDiffA}. The weighted average download times of
Method 1 is concentrated in the range of $[1.7,4.1]$ with mean value
2.877. Method 2 and Method 3 provide the upper bound and lower bound
for the minimum WADT. Fig.~\ref{fig:TimeA}, \ref{fig:TimeDiffA}, and
\ref{fig:RelTimeDiffA} show that these two bounds are very tight. In
this case, the lower bound is almost always the same as the optimal
solution since the distribution curves of Method 1 and Method 3
perfectly match. The distribution curve of Method 2 is slightly
different from that of Method 1, which means that the upper bound is
only slightly larger than the optimal solution for most of the
experiments. In this simulation, the bandwidth of the server is 10,
which is not much larger than the maximum of the upload bandwidths.
That is why the upper bound is still slightly different from the
optimal solution. The distribution of Method 5 shows that in most of
the experiments, the WADT of Method 5 is close to the optimal
solution, however, it is sometimes much larger than the optimal
solution. This is because the peer placement in Method 5 is usually
very good but sometimes bad. Note that it is NP complete complex to
find the best peer placement. In order to improve Method 5 without
increasing the complexity, we need provide a better but still
low-complex peer placement algorithm. The performance of Method 4 is
much worse than the performance of Method 5 because the peer
placement algorithm is Method 4 is much worse that in Method 5.

\begin{figure}
  \centering
  \includegraphics[width=0.5\textwidth]{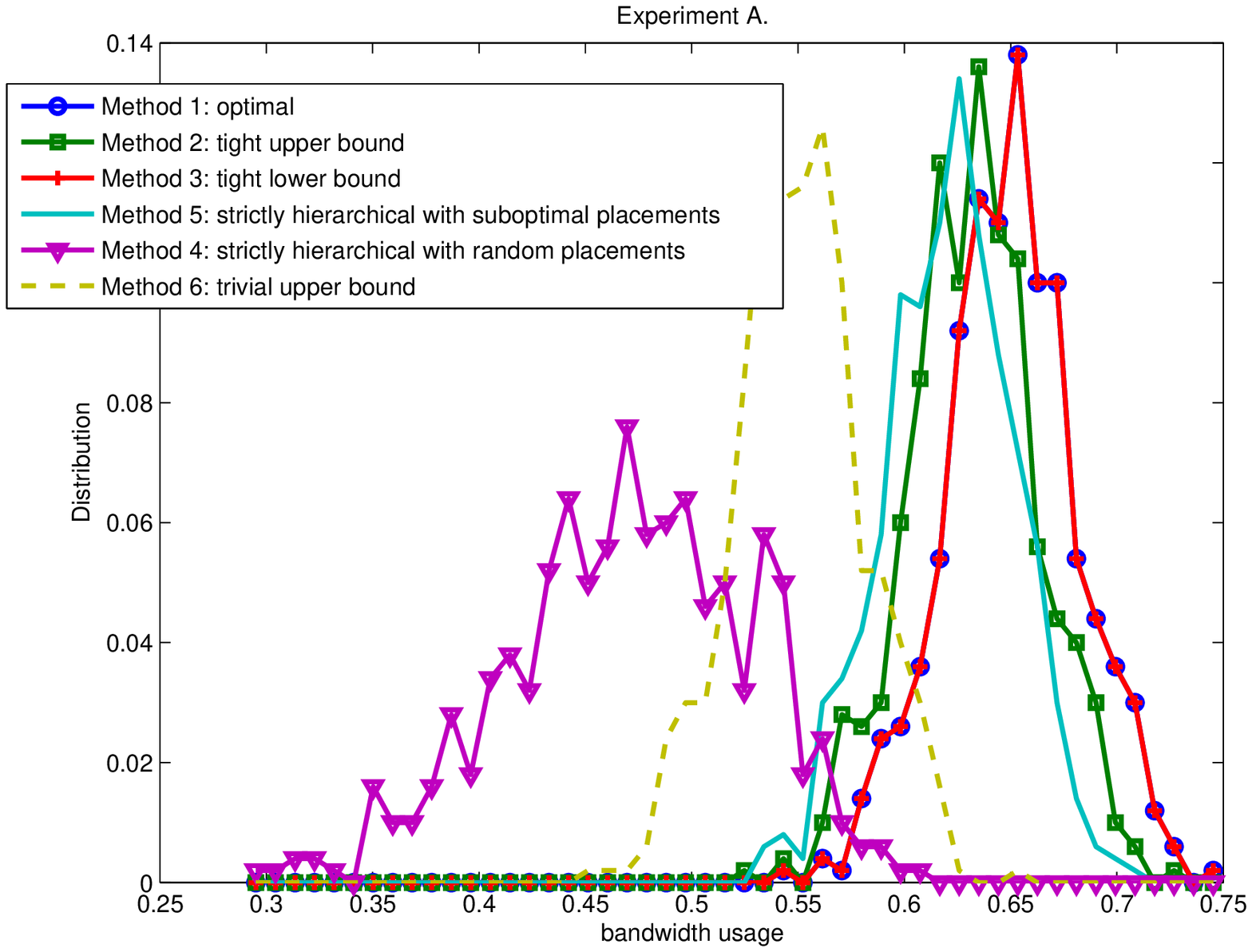}
  \caption{Distribution of the bandwidth usage for different
  methods.}\label{fig:ThroughputA}
\end{figure}

The bandwidth usage is the ratio of the total transmission rate and
the total download bandwidth of all the peers in the network. For a
fixed WADT, it is clear that the lower bandwidth usage the better.
However, it is a trade off to decrease the bandwidth usage and to
decrease the WADT, which is also verified in
Fig.~\ref{fig:ThroughputA}. Fig.~\ref{fig:ThroughputA} shows the
distribution of the bandwidth usage of different methods. The
distribution of Method 1, the optimal solution, is exactly the same
as the distribution of Method 3, the lower bound. The bandwidth
usage of Method 2, the upper bound, is slightly less than that of
Method 1 and the bandwidth usage of Method 5 is slightly less than
that of Method 2. It is verified that the method with smaller WADT
always has higher bandwidth usage. The mean values of the weighted
average download times and the bandwidth usage of different methods
are listed in Table \ref{table:A}.

\begin{table}
\begin{center}
\caption{The mean values of the weighted average download times
(W.A.D.T.), the normalized weighted average download times
(N.W.A.D.T.)and the bandwidth usage (B.U.)of different methods for
experiment A. }\label{table:A}

\begin{tabular}{|c|c|c|c|}

  \hline
  Method & W.A.D.T. & N.W.A.D.T. & B.U. \\
\hline
  7 & 2.468 & 0.854 & 1.000 \\
  \hline
  3  & 2.877 & 1.000 & 0.650 \\
  \hline
  1 & 2.877 & 1.000 & 0.650 \\
  \hline
  2  & 2.947 & 1.025 & 0.633 \\
  \hline
  5  & 3.461 & 1.200 & 0.621 \\
  \hline
  4 & 5.338 & 1.849 & 0.471 \\
  \hline
  6  & 6.361 & 2.194 & 0.551 \\
  \hline
\end{tabular}
\end{center}
\end{table}

\begin{figure}
  \centering
  \includegraphics[width=0.5\textwidth]{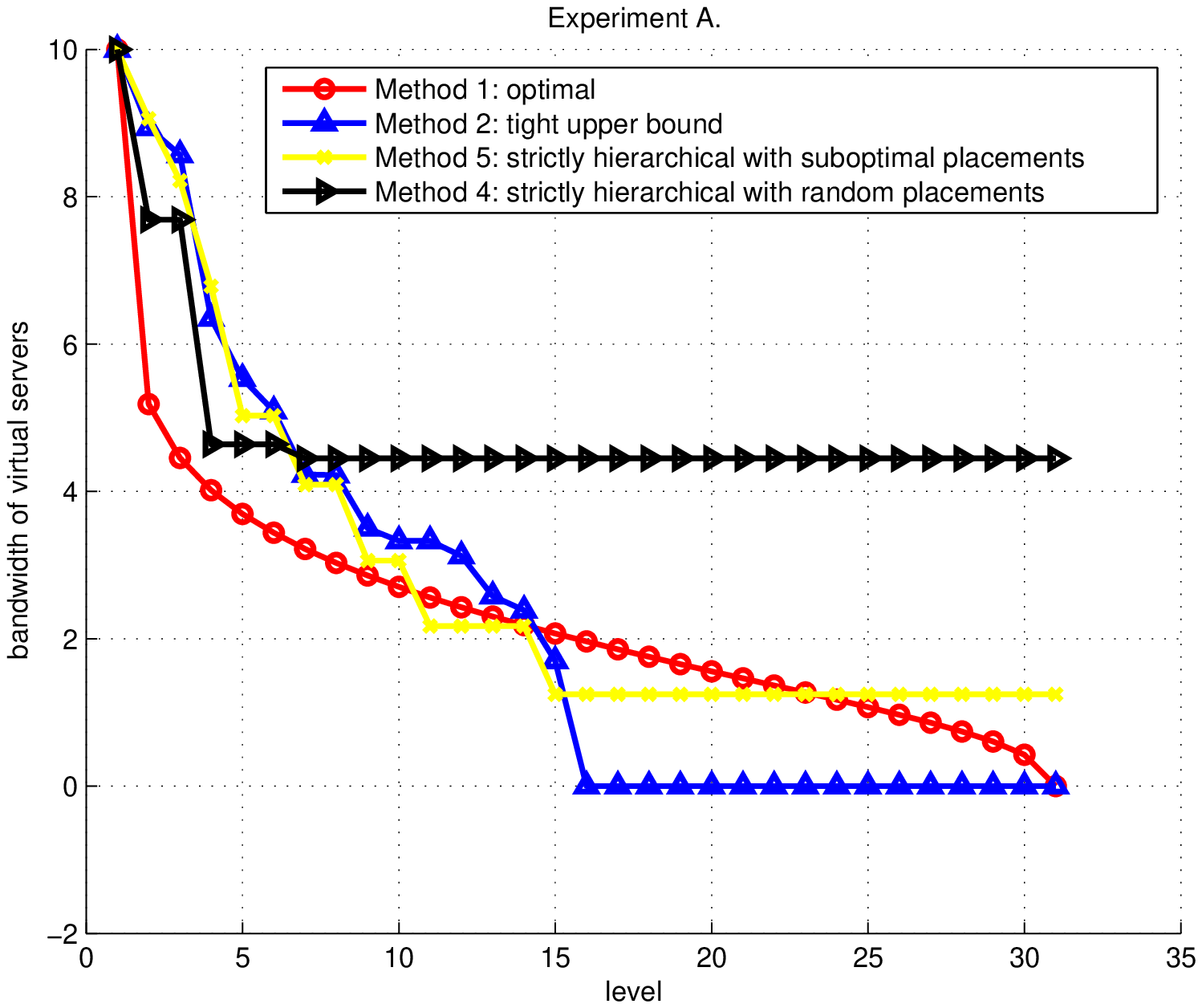}
  \caption{Bandwidth of virtual servers for different
  methods.}\label{fig:ServerA}
\end{figure}

For one typical experiment, the plots of the virtual server
bandwidths versus the levels for different methods are shown in
Fig.~\ref{fig:ServerA}. For Method 1, the number of the levels $K$
is manually chosen. In this experiment,$K$ is 30. One can see that
the bandwidths of the virtual servers from level 5 to level 30 are
linearly decreasing to 0 for Method 1. The bandwidth of the virtual
server generated by the last level is almost 0. In other words, all
the upload bandwidths of the server and the peers are fully used.
This is the reason why the performance of Method 3, the tight lower
bound, is almost the same as that of Method 1. For Method 2, the
number of the levels needed is automatically solved. It is 15 in
this experiment.

\subsection{A large network simulation}

In this simulation, the bandwidth of the server is 50 and there are
4000 peers in the network. The download bandwidths $d_i$,
$i=1,\cdots,4000$, are i.i.d. with uniform distribution over
$[0.01,1.99]$. The upload bandwidth $u_i$ is uniformly distributed
over $[0.1d_i,d_i]$. The weights for all the peers are equal and
normalized to be $1/N = 1/4000$, and so $\sum_i w_i = 1$.

\begin{figure}
  \centering
  \includegraphics[width=0.5\textwidth]{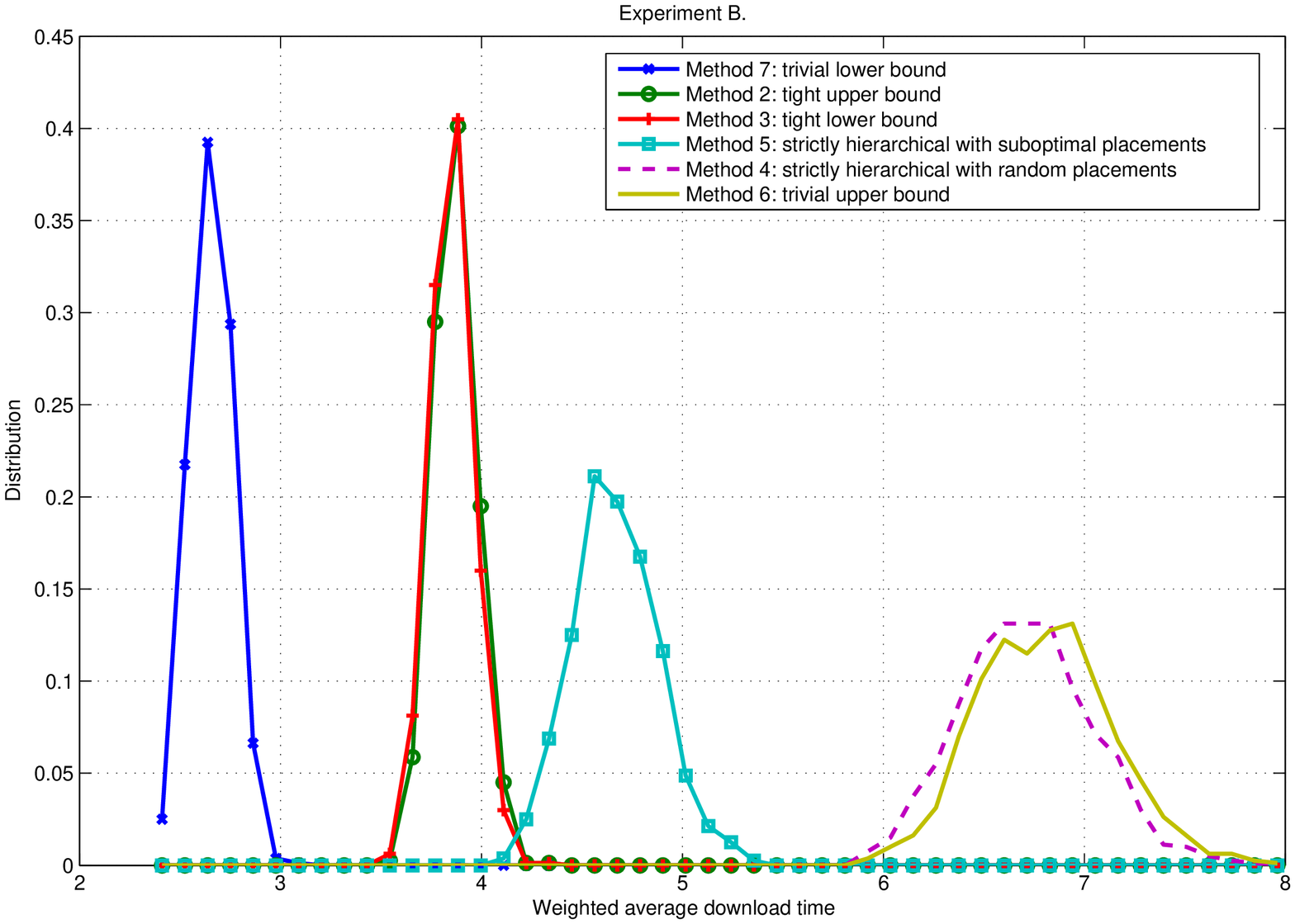}
  \caption{Distribution of the WADT for different
  methods.}\label{fig:TimeB}
\end{figure}

\begin{figure}
  \centering
  \includegraphics[width=0.5\textwidth]{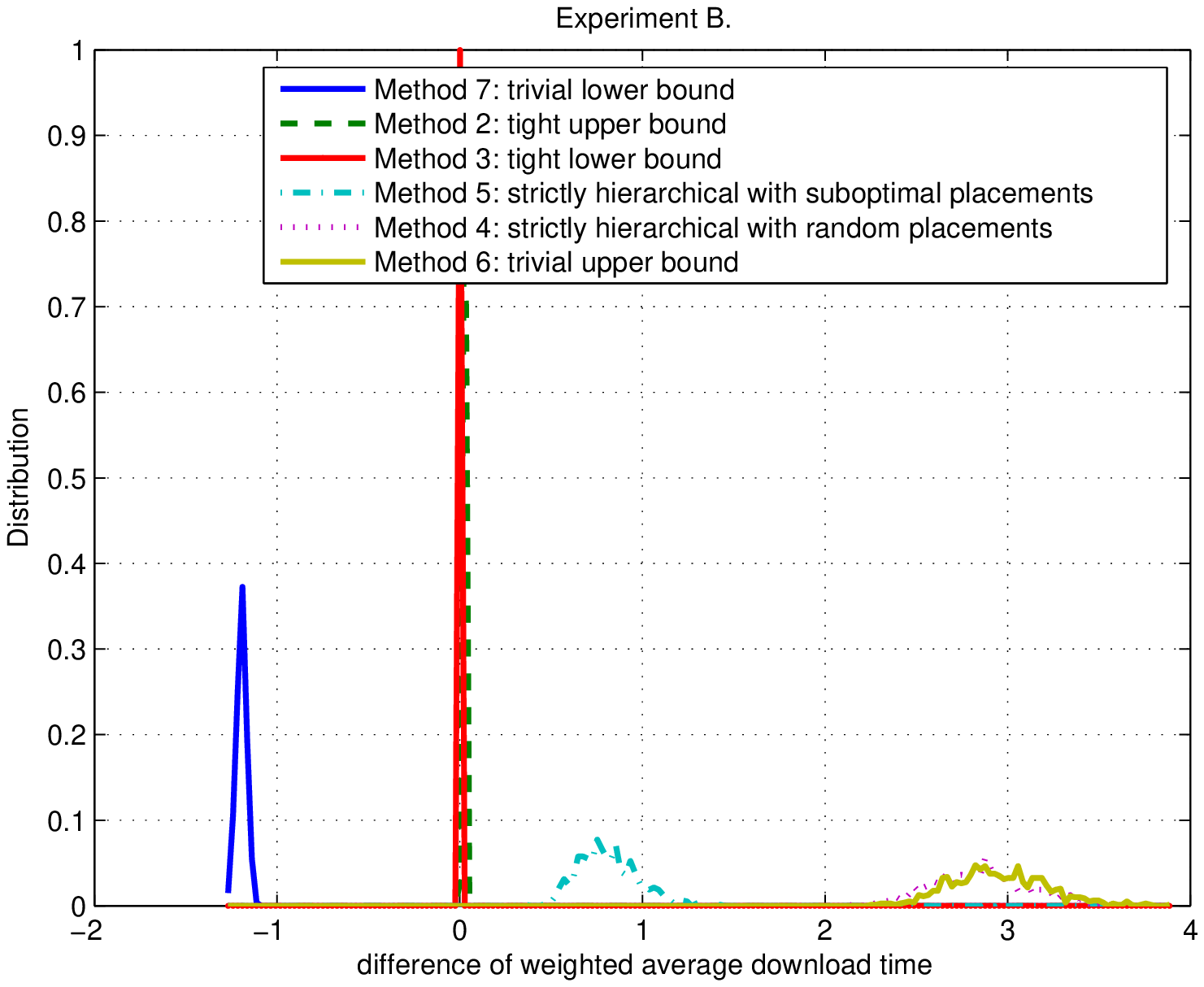}
  \caption{Distribution of the differences of the WADT between Method 3 and other methods.}\label{fig:TimeDiffB}
\end{figure}

\begin{figure}
  \centering
  \includegraphics[width=0.5\textwidth]{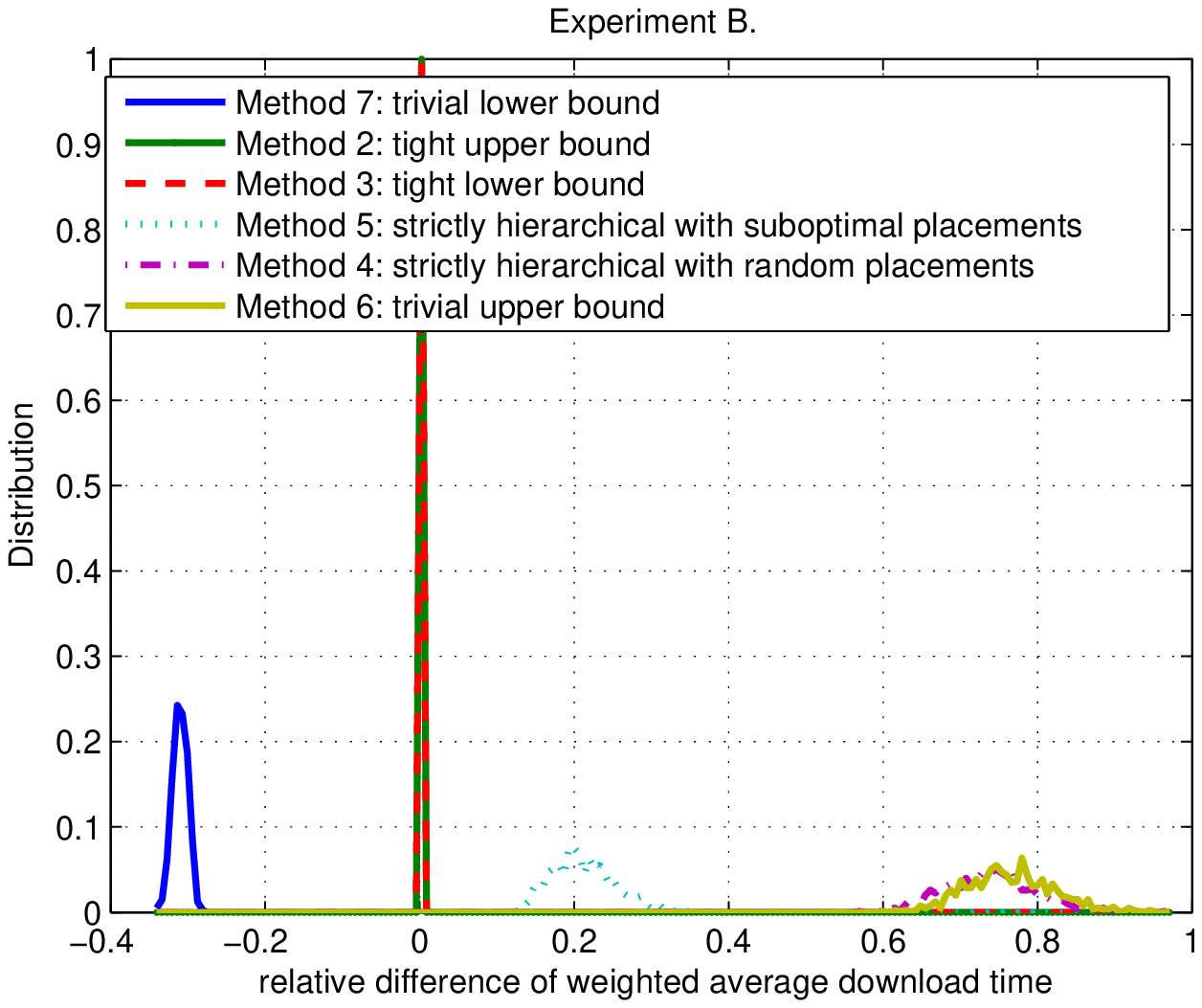}
  \caption{Distribution of the relative difference of the WADT between Method 3 and other methods.}\label{fig:RelTimeDiffB}
\end{figure}

The distribution of the weighted average download times of 800
experiments are shown in Fig.~\ref{fig:TimeB}. The distribution of
the difference and the relative difference between Method 3 and
other methods are shown in Fig.~\ref{fig:TimeDiffB} and
Fig.~\ref{fig:RelTimeDiffB}. These figures show that the lower bound
and the upper bound provided by Method 3 and Method 2 are very
close. The performance of Method 1 should be between these two
bounds, although we don't simulate Method 1 in this case. In this
simulation, the bandwidth of the server is 50, which is much larger
than the maximum of the upload bandwidths. That is why $S \simeq S -
\max_{i}(u_i)$ and the upper bound is almost the same as the lower
bound. The performance of Method 5 is worse than that of Method 2
but still a lot better that of Method 6 and 7.

\begin{figure}
  \centering
  \includegraphics[width=0.5\textwidth]{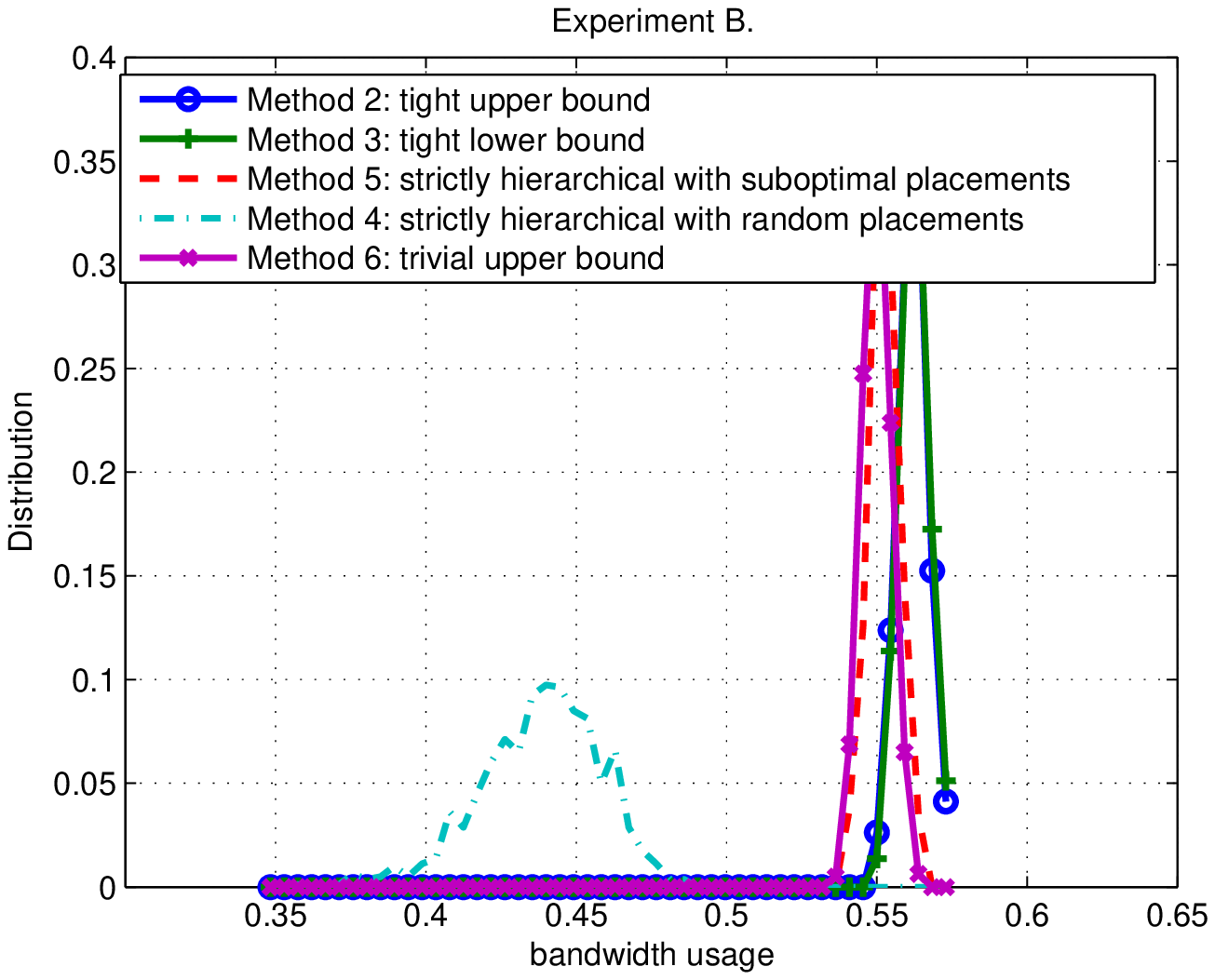}
  \caption{Distribution of the bandwidth usage for different
  methods.}\label{fig:ThroughputB}
\end{figure}

The bandwidth usages of these methods are shown in
Fig.~\ref{fig:ThroughputB}. The distribution of Method 2 is almost
the same as the distribution of Method 3, which verifies the almost
same performance of Method 2 and 3. The mean values of the weighted
average download times and the bandwidth usage of different methods
are listed in Table \ref{table:B}. Combining the simulation results,
the comparison of these 7 methods in different criteria are listed
in Table \ref{table:compare}.

\begin{table}
\begin{center}
\caption{The mean values of the weighted average download times
(W.A.D.T.), the normalized weighted average download times
(N.W.A.D.T.)and the bandwidth usage (B.U.)of different methods for
experiment B. }\label{table:B}

\begin{tabular}{|c|c|c|c|}

  \hline
  Method & W.A.D.T. & N.W.A.D.T. & B.U. \\
\hline
  7 & 2.659 & 0.6898 & 1.000 \\
  \hline
  3  & 3.854 & 1.000 & 0.562 \\
  \hline
  1 & - & - & - \\
  \hline
  2  & 3.870 & 1.041 & 0.562 \\
  \hline
  5  & 4.666 & 1.210 & 0.552 \\
  \hline
  4 & 6.709 & 1.740 & 0.438 \\
  \hline
  6  & 6.804 & 1.765 & 0.550 \\
  \hline
\end{tabular}
\end{center}
\end{table}

\begin{table}
\begin{center}
\caption{Comparison of the methods in the criteria of the weighted
average download time (W.A.D.T.), the bandwidth usage (B.U.), the
complexity of the computation in the server (S. Comp.) and the
complexity of the algorithms in peers (P. Comp.)
}\label{table:compare}

\begin{tabular}{|c|c|c|c|c|}

  \hline
  Method & W.A.D.T. & B.U. & S. Comp. & P. Comp. \\
\hline

  1 & Optimal & Highest & $O(K^3N^3)$ & High \\
  \hline
  2  & \scriptsize{Almost Opt.} & \scriptsize{Very High} & $O(N)$ & Low \\
  \hline
  5  & Good & High & $O(N)$ & Low \\
  \hline
  4 & Bad & Low & $O(N)$ & Low\\
  \hline
  6  & Bad & Low & $O(N)$ & Low \\
  \hline
\end{tabular}
\end{center}
\end{table}

\section{Conclusions}
\label{sec:conclusions}

This paper proposes an analytical framework for peer-to-peer (P2P)
networks and introduces schemes for building P2P networks to
approach the minimum weighted average download time (WADT). In the
considered P2P framework, the server, which has the information of
all the download bandwidths and upload bandwidths of the peers,
minimizes the weighted average download time by determining the
optimal transmission rate from the server to the peers and from the
peers to the other peers.

This paper first defines the static P2P network, the hierarchical
P2P network and the strictly hierarchical P2P network and studies
the graph structures of these P2P networks. The main result is that
any static P2P network can be decomposed into an equivalent network
of sub-peers that is strictly hierarchical. Therefore, convex
optimization can minimize the WADT for P2P networks by equivalently
minimizing the WADT for strictly hierarchical networks of sub-peers.
This paper then gives an achievable upper bound for minimizing WADT
by constructing a hierarchical P2P network, and lower bound by
weakening the constraints of the convex problem. Both the upper
bound and the lower bound are very tight.

The strictly hierarchical P2P network is practical for protocol
design because peer selection algorithms and chunk selection
algorithms can be locally designed level by level instead of
globally designed. Minimizing the WADT for strictly hierarchical
networks is a 0-1 convex optimization problem. However, if we have
assigned all peers each to a level, then the global bandwidth
allocation problem decomposes into local bandwidth allocation
problems at each level, which have water-filling solutions. Several
suboptimal peer assignment algorithms are provided and simulated.

\end{document}